 \documentclass[authoryear,preprint,review,12pt]{elsarticle}

\usepackage{amsmath}
\usepackage{array}
\usepackage{ulem}
\usepackage{xcolor}

\journal{Journal of Hydrology}

\begin{document}

\begin{frontmatter}


\title{Informed attribution of flood changes to decadal variation of atmospheric, catchment and river drivers in Upper Austria}



\author[label2]{Miriam Bertola \corref{cor1}}
\author[label3]{Alberto Viglione}
\author[label2]{G\"unter Bl\"oschl}

\address[label2]{Institute of Hydraulic Engineering and Water Resources Management, Vienna University of Technology, Karlsplatz 13, 1040, Vienna, Austria}
\address[label3]{Department of Environmental Engineering, Land and Infrastructure, Polytechnic University of Turin, Corso Duca degli Abruzzi 24, 10129 Torino, Italy}

\cortext[cor1]{Corresponding author: bertola@hydro.tuwien.ac.at}

\begin{abstract}
Flood changes may be attributed to drivers of change that belong to three main classes: atmospheric, catchment and river system drivers.
In this work, we propose a data-based attribution approach for selecting which driver best relates to variations in time of the flood frequency curve.
The flood peaks are assumed to follow a Gumbel distribution, whose location parameter changes in time as a function of the decadal variations of one of the following alternative covariates: annual and extreme precipitation for different durations, an agricultural land-use intensification index, and reservoir construction in the catchment, quantified by an index. The parameters of this attribution model are estimated by Bayesian inference.
Prior information on one of these parameters, the elasticity of flood peaks to the respective driver, is taken from the existing literature to increase the robustness of the method to spurious correlations between flood and covariate time series.
Therefore, the attribution model is informed in two ways: by the use of covariates, representing the drivers of change, and by the priors, representing the hydrological understanding of how these covariates influence floods.
The Watanabe-Akaike information criterion is used to compare models involving alternative covariates.
We apply the approach to 96 catchments in Upper Austria, where positive flood peak trends have been observed in the past 50 years.
Results show that, in Upper Austria, one or seven day extreme precipitation is usually a better covariate for variations of the flood frequency curve than precipitation at longer time scales.
Agricultural land-use intensification rarely is the best covariate, and the reservoir index never is, suggesting that catchment and river drivers are less important than atmospheric ones.
Not all the positive flood trends correspond to a significant correlation between floods and the covariates, suggesting that other drivers or other flood-driver relations should be considered to attribute flood trends in Upper Austria.
\end{abstract}


\end{frontmatter}



\section{Introduction}
\label{sec1:intro}
In recent years, a large number of major floods occurred, triggering many studies to focus on flood trend detection at local and regional scale \citep[see e.g.][for an European overview]{Mudelsee2003, Petrow2009, Bloschl2017, Mangini2018}. 
Despite trends in flood regime are detected in numerous studies, the identification of their driving processes and causal mechanisms is still far from being properly addressed \citep{Merz2012}. 
Understanding the reasons why the detected flood changes occurred (i.e. flood change attribution) is a complex task, since different processes, influencing flood magnitude, frequency and timing, can act in parallel and interact in different ways across spatial and temporal scales \citep[][]{Bloschl2007}. 
According to \citet{Pinter2006}, \citet{Merz2012} and \citet{Hall2014}, potential drivers of flood regime change belong to three groups: atmospheric, catchment and river system drivers.

The Atmospheric driver includes the meteorological forcing of the system (e.g. total precipitation, precipitation intensity/duration, temperature, snow cover/melt and radiation) whose changes can be related to both natural climate variability and anthropogenic climate change. They usually occur at large spatial scales, affecting flood regime consistently within a region, with gradual changes in time of the mean or the variance of peak discharges \citep[][]{Mudelsee2003, Bloschl2007, Petrow2009, Renard2014}.

The Catchment driver includes runoff generation and concentration processes, which are quantified, for instance, by the infiltration capacity or the runoff coefficient. They are susceptible to land-cover and land-use changes (e.g. urbanization, deforestation, change in agricultural practices) and are likely to occur gradually in time, usually with diminishing effects with increasing catchment area \citep[][]{Bloschl2007, OConnell2007, Rogger2017, Alaoui2018}.

The River System driver includes flood wave propagation processes into the river network. River training and hydraulic structures produce modifications of river morphology, roughness, water levels, discharge and inundated area, resulting typically in step changes in the time series of flood discharge peaks. 
Usually, these changes occur in proximity (e.g. flood flow acceleration and channel incision) or downstream (e.g. loss of floodplain storage) of the river modification, e.g. downstream of reservoirs or downstream urban areas, where structural flood protection measures are developed \citep{Graf2006, Pinter2006, Volpi2018}.

In the past, as pointed out by \citet{Merz2012}, the attribution of flood changes has been mainly done through qualitative reasoning, suggesting relationships with changes in climate variables (e.g. precipitation or circulation patterns) or anthropogenic impacts (e.g. river training, dam construction or land-use change), and citing literature to support these hypotheses. 
Recently, however, in several studies the detected flood changes are quantitatively related to one or, more rarely, to more than one of the potential drivers. 
This has been done essentially in two different ways: the data-based and the simulation-based approach. 

The data-based approach consists in identifying the relationship between drivers and floods from data only, in a statistical way.
For example, studies exist that analyze the correlation and geographic cohesion between flood characteristics and large-scale climate indices \citep{Archfield2016} or the long-range dependencies of precipitation and discharge \citep{Szolgayova2014} and their spatial and temporal co-evolution \citep{Perdigao2014}.
Many studies use the so called "non-stationary flood frequency analysis" to improve the reliability of flood quantile estimation by relating the parameters of flood frequency distributions to covariates, such as large-scale climate indices or large-scale atmospheric or oceanic fields \citep[i.e. climate-informed frequency analysis, see e.g.][]{Renard2014, Steirou2018}, extreme precipitation \citep{Villarini2009, Prosdocimi2014}, annual precipitation \citep{Sraj2016}, reservoir indices \citep{Lopez2013, Silva2017}, population measures \citep{Villarini2009}, etc. 
The advantage of the data-based approach, when compared to other methods, is that, due to its relative simplicity, it is easily applicable to many sites, at the regional or even continental scale.
Its drawback is that it identifies correlations between covariates and flood dynamics, usually without investigating whether the magnitude of these correlations are consistent with what process understanding would suggest.

Cause-effect mechanisms are instead included in the simulation-based approach, which consists in reproducing the observed flood changes by introducing, in hydrological models, changes in the potential driver(s) and observing the effects on the simulated hydrograph characteristics \citep{Merz2012}. 
Several simulation-based studies analyze the effects of extensive river training on flood regime \citep[see e.g.]{Lammersen2002, Vorogushyn2013, Skublics2016}. 
The effect of land-use changes (e.g. forestry management, agricultural practices and urbanization) on discharge is often investigated, in simulation-based studies, for specific catchments and flood events, under different land-management scenarios \citep[see e.g.][]{Niehoff2002, Bronstert2007, OConnell2007, Salazar2012}. 
The advantage of the simulation-based approach is that process understanding is explicitly taken into account.
However, due to the complexity of the models, simulation-based methods are usually applied to single (or few) catchments at a time.

Clearly, it would be of interest to make use of the advantages of both approaches, when performing attribution studies.
\citet{Viglione2016}, propose a framework for attribution of flood changes, based on a regional analysis, that make use of process understanding in a data-based analysis.
They exploit information, obtained through rainfall-runoff modelling, on how different drivers should affect floods for catchments of different size.
The estimation of the relative contribution of the drivers is framed in Bayesian terms and the process-based information is quantified by prior knowledge about the scaling parameters of the regional model.

In this paper we also make use of knowledge accumulated in previous studies relating floods to dominant drivers, when performing attribution.
We use the same study region of \citet{Viglione2016}, where positive trends in flood peak series are observed, but differently from them, who focus on attribution at the regional level, we are interested in the attribution at the local (site-specific) scale. 
We apply the non-stationary flood frequency method, here called "driver-informed" flood frequency method \citep[consistently with][]{Steirou2018}, to 96 sites in Upper Austria, using local (rather than regional) covariates on atmospheric, catchment and river system drivers.
Differently from \citet{Viglione2016}, we allow the drivers to act in opposite directions when contributing to positive flood peak changes.
We use Bayesian inference for parameter estimation, with prior information on the connection between covariates and flood peaks taken from previous studies, both data-based and simulation-based ones.
The attribution is performed by comparing alternative models (with alternative covariates) using an information criterion that quantifies how well the flood frequency model fits the flood data (accounting for prior information) and penalize models that are too complex given the information available.
The attribution model is therefore informed in two ways: by the use of covariates, representing the drivers of change, and by the priors, representing the hydrological understanding of how these covariates influence floods.

Section 2 describes the driver-informed flood frequency model and the way attribution is performed. Section 3 describes the data used, including how information from the literature is translated into prior knowledge on the model parameters.
Section 4 reports the results of the analysis, investigating the sensitivity of the attribution results to different time-scales of the atmospheric driver and the dependency of the driver effects on the catchment area \citep[as hypothesized by][]{Hall2014, Viglione2016}.

\section{Methods}

\subsection{Flood Frequency analysis and alternative driver-informed models} 

For simplicity, we assume the maximum annual peak discharges to follow a two-parameter Gumbel distribution. 
Visual inspection of the data in Gumbel probability diagrams shows consistency with this assumption for most of the sites (note that the following procedure can be applied using more flexible distributions, i.e. with more parameters, without loss of generality).
The Gumbel cumulative distribution function is defined as:
\begin{equation}
G(z)= \exp \bigg\{ -\exp \bigg\{ -\frac{z-\mu}{\sigma} \bigg\} \bigg\}
\end{equation}
where $\mu$ and $\sigma$ are respectively the location and scale parameter of the distribution. These parameters are usually assumed invariant in time. 

In recent studies, climate variables have been used as covariates for the extreme value distribution parameters, which are therefore not constant in time. 
This approach is usually called "non-stationary" even if the resulting distribution can be considered non-stationary only if the covariates exhibit a deterministic change in time \citep[][]{Montanari2014, Serinaldi2015}. 

We use local covariates of the extreme value distribution parameters, representative for the three drivers of flood change (i.e. the atmospheric, catchment and river system processes) in the study region, and, similarly to the climate-informed statistics of \citet{Steirou2018}, we refer to this as driver-informed distribution/parameters. 

The following models are considered:
\begin{align}
&G_0)   &  \mu &= \mu_0,             &  \sigma&=\sigma_0\\
&G_1)   &  \log(\mu)&=a+b \log(X),   &  \sigma&=\sigma_0\\
&G_2)   &  \log(\mu)&=a+bX,          &  \sigma&=\sigma_0
\end{align}
where $X$ is a general covariate (e.g. one of the drivers) and $a$ and $b$ are regression parameters to be estimated locally. 
The location parameter $\mu$ only is conditioned on the external covariate, with two different dependence structures in model $G_1$ and $G_2$. 
Practically speaking, they introduce one additional parameter to be estimated, compared to the time-invariant Gumbel distribution $G_0$. 
The parameters are estimated by fitting the alternative models to flood data with Bayesian inference through a Markov Chain Monte Carlo approach. 
The R package \textit{rStan} \citep{Carpenter2017} is used to perform the MCMC inference. rStan makes use of Hamiltonian Monte Carlo sampling, which speeds up convergence and parameter exploration by using the gradient of the log posterior \citep{StanManual}. 
For each inference, we generate 4 chains of length $N_{sim}=10000$, each starting from different parameter values, and check for their convergence.

One advantage of the Bayesian framework is the possibility to take into account additional prior belief (e.g. expert knowledge) or external a priori information about the parameters in their estimation. 
Herein, we set informative priors on the parameter $b$, based on the results of published studies (see Section 3.4), in order to limit the possibility for spurious correlations to bias the attribution.
In model $G1$ the parameter $b$ is defined as:
\begin{equation}
b = \frac{X}{\mu} \cdot \frac{d\mu}{dX}
\end{equation}
and represents the percentage change of the location parameter of the distribution of annual maxima, following a 1\% change in the covariate $X$. 
In other words, the parameter $b$ represents the elasticity of (the location parameter of) flood peaks with respect to the covariate, similarly to the temporal sensitivity coefficient of flood to precipitation defined in \citet{Perdigao2014}. 
In model $G2$ instead, the parameter $b$ is defined as:
\begin{equation}
b = \frac{1}{\mu} \cdot \frac{d\mu}{dX}
\end{equation}
It represents the relative change occurring in the location parameter of the distribution of annual maxima, following a unit change in the covariate. 

\subsection{Model selection and flood change attribution}
The Widely Applicable or Watanabe-Akaike Information Criterion (WAIC) is used in this study for model comparison and selection. 
Its measure represents a trade-off between goodness of fit and model complexity.
The WAIC, originally proposed by \citet{Watanabe2010}, is one of the Bayesian alternatives of the Akaike Information Criterion (AIC) \citep{akaike1973information}. 
It estimates the out-of-sample predictive accuracy ($elppd$) by subtracting, to the computed log pointwise posterior predictive density ($lppd$), a penalty for the complexity of the model expressed in terms of effective number of parameters ($p_{WAIC}$) \citep{Gelman2014}. 
We evaluate the WAIC as defined in \citet{Gelman2014} and in \citet{Vehtari2017}:
\begin{equation}
WAIC = -2 \cdot \widehat{ellpd}_{WAIC} = -2 \cdot (lppd-p_{WAIC})
\end{equation}
Where the multiplication factor -2 scales the expression, making it comparable with AIC and other measures of deviance. 
The R package \textit{loo} is used for the calculations.

\section{Study area and drivers of flood change}
\begin{figure}
 \center\includegraphics[width=32pc]{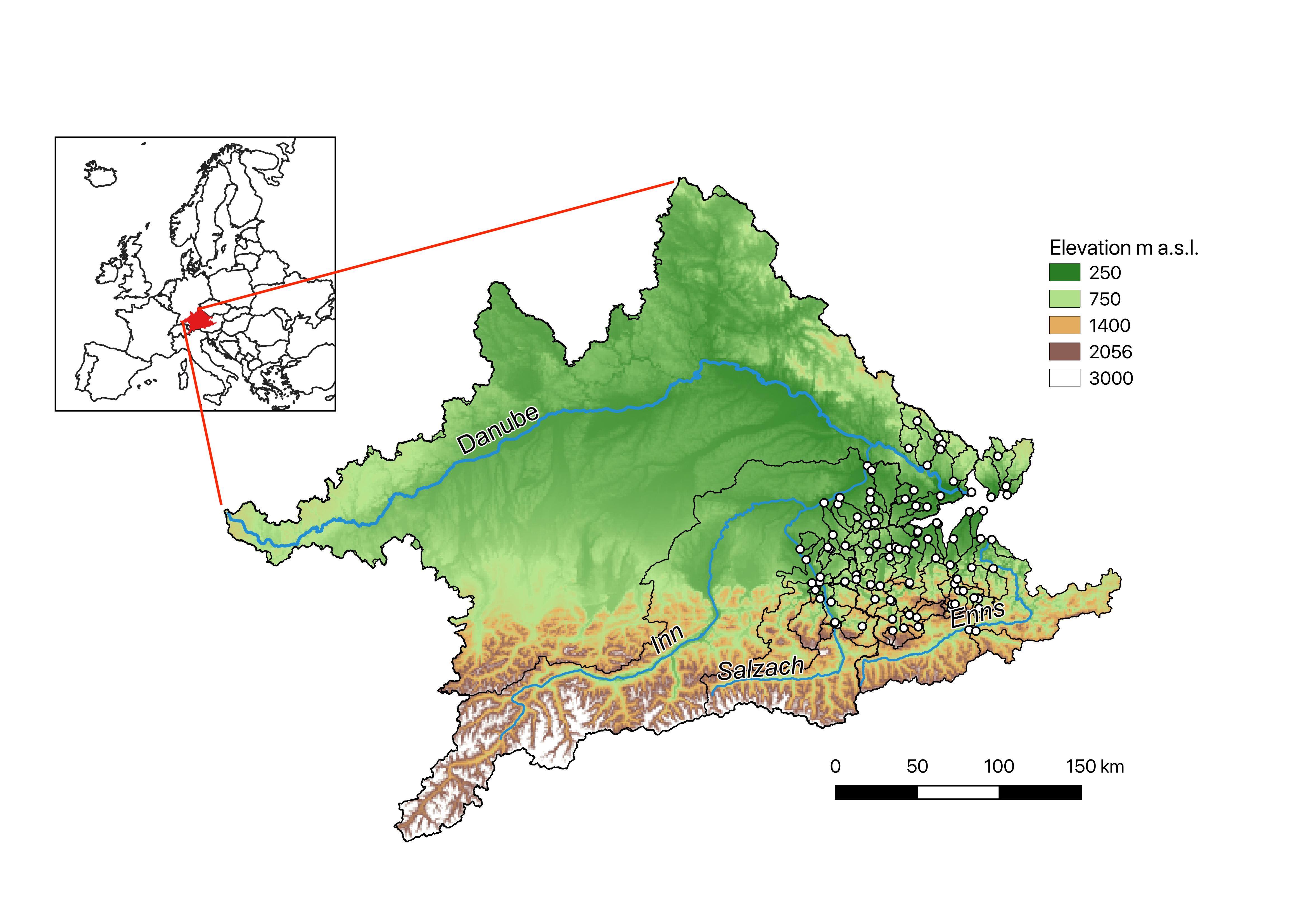}
 \caption{Study region. Location and elevation of the 96 catchments, with outlets in Upper Austria.}
 \label{fig1}
\end{figure}

As in \citet{Viglione2016}, the study area considered is Upper Austria, where annual maximum daily discharges (AM) for 96 river gauges (catchment areas ranging from 10 to 79500 $km^2$) are available with record lengths of at least 40 years after 1961. 
Figure \ref{fig1} shows the extension and the elevation of the considered catchments and Table \ref{tab1} contains percentiles of some catchment attributes.
 
\begin{table}
\small
 \begin{tabular}{llllll}
  \hline
  Percentile:                            & 0\%   & 25\%  & 50\%  & 75\%  & 100\%   \\ \hline
  Catchment area (km2):                  & 10.5  & 68.6  & 159.4 & 428.2 & 79490.1 \\
  Elevation of the outlet (m a.s.l.):    & 246.7 & 357.0 & 442.1 & 504.1 & 763.5   \\
  Mean annual flow (m$^3$/s):            & 0.2   & 1.6   & 3.9   & 10.9  & 1583.0  \\
  Mean annual flood (m$^3$/s):           & 6.2   & 24.5  & 46.7  & 138.1 & 4415.3  \\
  Length of the flood series (years):    & 40    & 54    & 64    & 96    & 182     \\ \hline
\end{tabular}
\caption{Percentiles of catchment attributes (catchment area, outlet elevation, mean annual flow, mean annual flood and length of records) over the 96 considered catchments}
\label{tab1}
\end{table}

\begin{figure}
 \center\includegraphics[width=32pc]{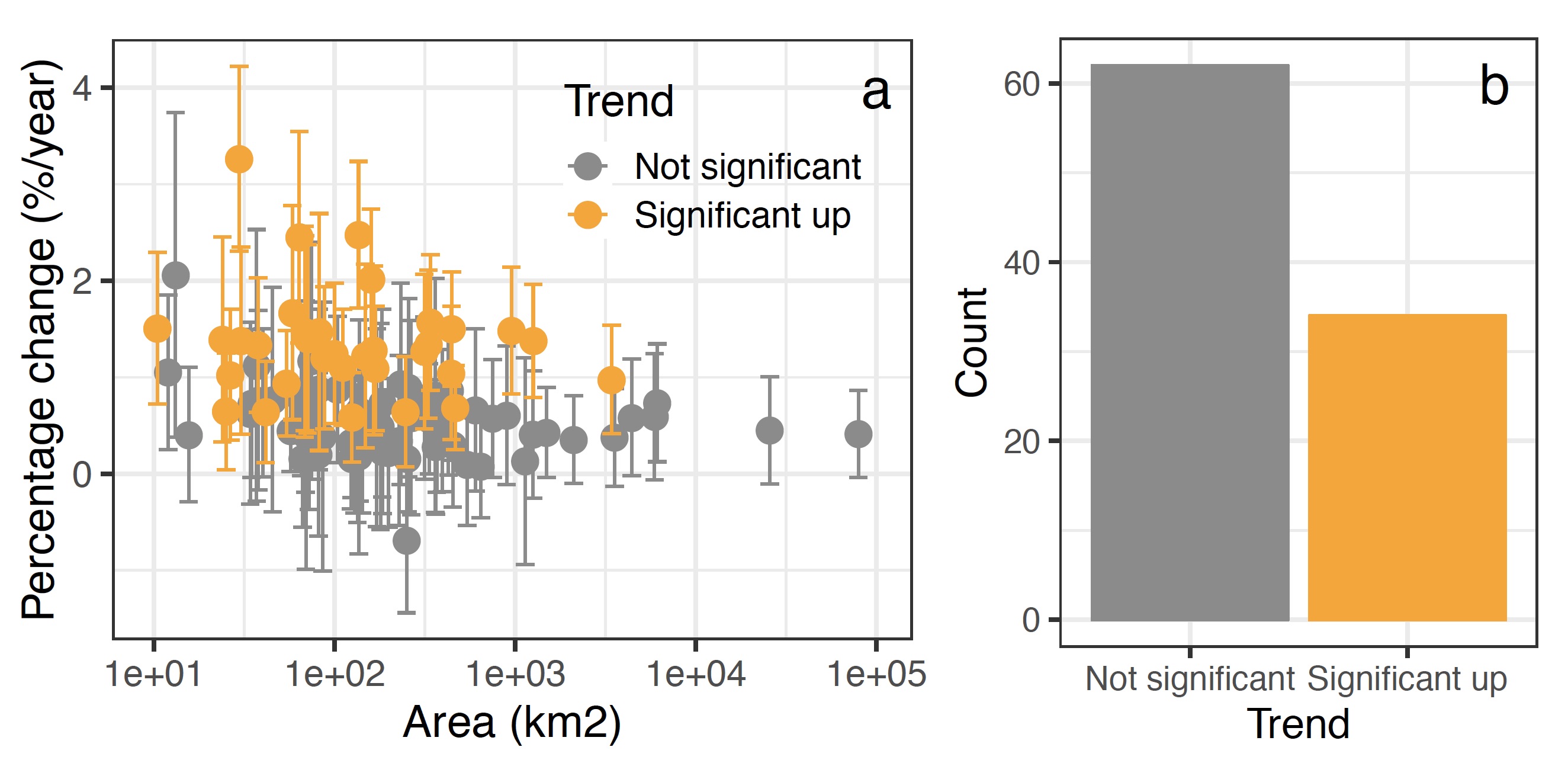}
 \caption{Detected trends (in $\% \: {year}^{-1}$) in the annual maximum discharge with 95\% confidence intervals, as a function of catchment area \citep[as in][]{Viglione2016} (panel a). Significant upward trends (based on Mann-Kendall test at 5\% significance level) are represented in orange. Panel b shows the occurrence of significant upward vs not significant trends in the region.}
 \label{fig2}
 \end{figure}

In the considered region, clear evidences of positive trends in flood peaks have been detected in previous studies \citep{Bloschl2011,Bloschl2012,Viglione2016}. 
Figure \ref{fig2} (panel a) shows the trends in the logarithm of the flood peaks (this is equivalent to the percentage change in time), together with their 95\% confidence intervals, resulting from a simple least square linear regression, taking 1961 as a common starting year of the AM series. 
Mostly positive trends are detected, with magnitude between -1 and 3.5 \% change per year.
A common Mann-Kendall test with 5\% significance is performed to identify significant trends (shown in orange in the figure).
Panel b shows that more than one third of the catchments in the region has a positive significant trend over time.
 
In this study, instead, we search for relationships between flood temporal variations and the long term evolution of precipitation (atmospheric driver), land-use and agricultural intensification (catchment driver) and the construction of reservoirs (river system driver).
Table \ref{tab2} contains some statistics of the covariates (and related quantities) that we use, as possible drivers of flood change, in the driver-informed models $G_1$ and $G_2$.
\begin{table}
\footnotesize
\begin{tabular}{llllll}
\hline
Percentile:                                    & 0\%  & 25\% & 50\% & 75\% & 100\% \\ \hline
Mean annual precipitation (mm):                & 762.4  & 1081.2 & 1353.5 & 1641.6 & 2153.2  \\
30-day annual max. precipitation (mm):      & 164.7  & 218.4  & 257.4  & 308.5  & 413.7   \\
7-day annual max. precipitation (mm):       & 81.6   & 103.3  & 126.8  & 155.5  & 214.8   \\
1-day annual max. precipitation (mm):       & 35.0   & 44.1   & 51.6   & 61.9   & 82.2    \\
Crop area fraction (\%):                       & 0.0  & 1.5  & 4.7  & 14.2 & 91.6  \\ 
Mean maize yield in year 2000 (t/ha):      & 0.00 & 2.10 & 6.09 & 9.23 & 9.68  \\ 
Mean Land-use intensity Index (-):             & 0.00 & 0.01 & 0.03 & 0.13 & 0.83  \\
Reservoir capacity sums (10$^6$ m$^3$):                 & 0.0  & 0.0  & 0.0  & 0.0  & 1376.1\\ 
Mean Reservoir Index (-):                      & 0.00 & 0.00 & 0.00 & 0.00 & 0.05  \\ \hline 
\end{tabular}
\caption{Percentiles of the covariates and some covariate-related quantities, calculated over the 96 catchments}

\label{tab2}
\end{table}

\subsection{Long-term evolution of precipitation}
Daily precipitation records from 1961, averaged over each catchment, are obtained from the Spartacus gridded dataset of daily precipitation sum (spatial resolution 1x1 km) \citep{Hiebl2018}.
We extract extreme precipitation series (i.e. 30-day, 7-day and 1-day annual maximum precipitation), commonly used as covariates in the literature \citep[e.g.][]{Prosdocimi2014,Villarini2009}, and annual total precipitation (see Table \ref{tab2}). 
This latter is the preferred predictor of flood frequency changes in some studies \citep[e.g.][]{Perdigao2014,Sivapalan2015,Sraj2016} and is here considered as a proxy of the antecedent soil moisture condition before a flood event \citep{Mediero2014} as well as of the event precipitation.

In this study, we consider the decadal variation of the mean annual maximum precipitation for different durations and the annual total precipitation as potential drivers of the decadal variation of the annual flood peak discharges. Therefore, as we are interested in this long term evolution rather than in the year-to-year variability, we smooth the precipitation series with the locally weighted polynomial regression LOESS \citep{Cleveland1979} using the R function \textit{loess}. 
The subset of data over which the local polynomial regression is performed is 10 years (i.e. 10 data-points of the series) and the degree of the local polynomials is set equal to 0. 
This is equivalent to a constant local fitting and turns LOESS into a weighted 10-years moving average. 
The weight function used for the local regression is the tri-cubic weight function. 
The locally weighted polynomial regression is used, rather than a common moving average, in order to preserve the original length of the series.

\subsection{Land-use change and intensification of field crop production}
We investigate the impact (at the catchment scale) on floods of modern agricultural management practices and heavy machineries, producing soil compaction and degradation \citep{VanDerPloeg1999,VanderPloeg2001,VanderPloeg2002,Niehoff2002,Pinter2006}.
With the exception of the mountainous catchments located mainly in the southern part of the region, agricultural areas cover significant portions of the catchments, with 290000 $ha$ (i.e. $\sim$ 25\% of the region area) of cropland in total over the region \citep{lko2016}.

A catchment-related land-use intensity index $LI$,  with a structure similar to the Reservoir Index, proposed by \citet{Lopez2013}, is built here. 
It is defined as:
\begin{equation}
LI=\sum_{i=1}^{N} \frac{A_{c,i}}{A_T} \cdot \frac{Y_i}{Y_{ref}}
\end{equation}
where $N$ refers to the number of sub-areas (i.e. the grid cells) contained into the catchment boundaries, $A_{c,i}$ is the cropland area, $Y_i$ is the yield in tons/ha, $A_T$ is the total catchment area and $Y_{ref}$ is the Reference yield. 

This land-use intensity index takes into account both the intensification of agricultural production \citep[represented by the ratio $Y_i$/$Y_{ref}$, similar to the $\tau$-factor in][as a proxy agricultural land use-intensity]{Dietrich2012}, and the land-use of the catchment (represented by the ratio $A_{c,i}$/$A_T$) with its potential change in time.

Cropland area $A_{c,i}$ is derived for each catchment from the globally available dataset of cropland and pasture areas for the year 2000, provided by \citet{Ramankutty2008} on a 5 min by 5 min latitude/longitude ($\sim$ 10 km by 10 km) grid. 
It combines agricultural inventory data with satellite-derived land cover data. 
We considered the ratio $A_{c,i}$/$A_T$ constant over time, since there are no substantial evidences of land-use changes over the period of interest in the region.
In other words, the changes of LI are, in this case, due to the intensification of the agricultural production only.

For what concerns yield data, we focus on the production of maize, which is the most important crop in Upper Austria \citep{lko2016}. 
Furthermore, \citet{Beven2008} list maize among the cropping systems associated with compaction and soil structural damage, due to the required practices (e.g. they keep bare soil surface) and type of operations, their timing (i.e. late harvested crops, requiring access to the soil during the wettest soil period, causing compaction, and leaving bare soil exposed to winter storms) and depth of cultivation \citep{Chamen2003}. 
Maize yield data for the year 2000 \citep[provided by][]{Monfreda2008} and its linear trend in time \citep[provided by][]{Ray2012} are globally available, in form of 5 min by 5 min latitude/longitude gridded data-sets. 
Time series of maize yield for each catchment are derived from spatial aggregation of the gridded information and by extrapolation of the linear trends over the period 1961-2014. 

The reference yield $Y_{ref}$, differently from \citet{Dietrich2012} where it represents the obtainable yield under standard and static agricultural management practices and varies with space, is here assumed to be a single value for the entire region, representative for its average maize production. 
It is calculated by averaging over time the field crop production data for maze in Upper Austria provided by \citet{stataustria} (in tons and hectares) and available for the period 1971-2017. 
The resulting $Y_{ref}$ is 8.72 ton/ha.  
See Table \ref{tab2} for statistics about the $LI$ in the region.

\subsection{Potential impact of reservoirs}
Within the 96 considered catchments, 21 reservoirs and the corresponding dams, are identified using the Global Reservoir and Dam GRanD database \citep[][]{grand2011}. 
Dam location, year of construction, capacity and drainage area of the reservoir are extracted from the GRanD database and used in this framework (see Table S1 in the Supplementary material for details). 
The potential impact of reservoirs on flood regime is here quantified using the Reservoir Index (RI) proposed by \citet{Lopez2013} and defined as follows:
\begin{equation}
RI=\sum_{i=1}^{N} \frac{A_i}{A_T} \cdot \frac{C_i}{C_T}
\end{equation}
Where $N$ is the number of reservoirs upstream of the gauge station, $A_i$ and $C_i$ are the catchment area and the capacity of each reservoir and $A_T$ and $C_T$ are the catchment area and the mean annual flow volume at the gauge station. 
The construction of a dam represents a step change in the $RI$. 
\citet{Lopez2013} find 0.25 to be $RI$ threshold value between low and high flow alteration.
See Table \ref{tab2} for statistics about the RI in the region.

\subsection{Driver-informed models and prior knowledge}
We use the drivers of change, described in section 3.1, 3.2 and 3.3, as covariates $X$ of the driver-informed models of section 2.2.
We adopt the model $G_1$ when investigating the effects on floods of the long-term evolution of precipitation (i.e. where $X$ is one of the smoothed precipitation series described in section 3.1, here generally indicated as $P$), otherwise we adopt model $G_2$, when investigating the effects of the agricultural soil degradation or reservoir (i.e. where $X$ is the $LI$ or $RI$). 
The alternative Gumbel distributions, with location parameter conditioned on the covariates are:
\begin{align}
&G_A)   &  \log(\mu)&=a_A+b_A \log(P),   &  \sigma&=\sigma_{0,A}\\
&G_C)   &  \log(\mu)&=a_C+b_C \cdot LI,  &  \sigma&=\sigma_{0,C}\\
&G_R)   &  \log(\mu)&=a_R+b_R \cdot RI,  &  \sigma&=\sigma_{0,R} 
\end{align}
This choice comes from the hypothesis that, when investigating the effects of the agricultural soil degradation or reservoir on floods, the actual magnitude of the covariate and its absolute variation is important, and not the relative change (e.g. an increase of 10\% of the cropland area may be not influential for floods if the initial cropland area is very small). This corresponds to the model structure $G_2$ and the related regression parameter $b$ as defined in Eq.6. 
On the contrary, when considering the atmospheric driver, we want the regression parameter $b$ to represent the elasticity of floods to precipitation. This is consistent with the temporal sensitivity coefficient of flood to precipitation of \citet{Perdigao2014} and corresponds to model $G_1$ and Eq.5. 
Note that the structure of the driver-informed models and the drivers/covariates considered are both assumptions that may be varied. With the proposed framework, we compare alternative models, that reflect/contain these assumptions for the considered region. Other models can be easily formulated to reflect other hypotheses. 

Informative a priori on the parameters $b_A$, $b_C$ and $b_R$ are retrieved from a selection of published studies, listed in Table \ref{tab3} (as for the model structure and the drivers, they are also part of the assumptions made). 
They evaluate the effects of the change in one of the drivers on the magnitude of flood peaks (i.e. they provide information on the value of the parameters $b$, as defined in Eq. 10, 11 and 12). 
The following paragraphs describe in detail the procedure followed to retrieve an estimate of the mean and the variance of their prior distribution, for each of the three drivers of change.

\paragraph{Atmospheric driver}
\citet{Perdigao2014} provide, in their Table 2, spatiotemporal sensitivity coefficients $\alpha$ and $\beta$ of floods to annual precipitation, together with 95\% confidence intervals, for Austria and its five hydroclimatic regions, obtained analyzing AM series of 804 catchments. 
The mean and standard deviation of the prior distribution of the parameter $b_A$, defined consistently with the sensitivity coefficient $\beta$ in the time domain, are taken respectively equal to 0.61 (value provided in the study for $\beta$) and 0.06 (obtained from its 95\% confidence bounds with the assumption of normality).
We adopt these values as moments of the prior normal distribution of $b_A$ when the covariate is annual precipitation \citep[as in][]{Perdigao2014}, but also when the covariate is one of the extreme precipitation series. 
In these latter cases, in order to reflect the additional uncertainty related to this choice, we arbitrarily increase the standard deviation to three times the one in \citet{Perdigao2014} (i.e. 0.18). 

\paragraph{Catchment driver}
The impact of agricultural soil compaction on flood peaks at the catchment scale is still underdeveloped in the scientific literature \citep{Rogger2017} and it is not possible to directly retrieve a priori on the regression parameter $b_C$, as defined in this framework. 
For this reason, we assume that the available prior information related to land-use change can be transferred and used when analyzing the effect of land-use intensification on floods.
\citet{Fraser2013} present an application of metamodeling that upscales physics-based model predictions to make catchment scale predictions of land-management change impacts on peak flows. 
They consider four land-management scenarios, involving changes of land-use between 3 and 30\% of catchment area in one catchment (river Hodder at Footholme in north-west England, 25.3 km$^2$), whose size and agricultural nature is consistent with most of the catchments in this study.
For each scenario they provide, in their Table 4, the minimum, median and maximum reduction of the mean catchment peak flow predicted with two different modelling approaches. 
The mean of the prior distribution of $b_C$ is obtained dividing the predicted mean catchment peak flow reductions (we consider the values in the column "median") by the imposed fraction of area under land-use change of the corresponding scenario, and finally averaging over the scenarios. 
The resulting mean of the distribution of $b_C$ is 0.13. 
The predicted minimum and maximum reductions of the mean peak flow are also divided by the corresponding land-use change and averaged over the scenarios, obtaining a minimum and maximum predicted value for $b_C$. 
We treat these latter as 95\% confidence bounds of reduction of the mean catchment peak flow, from which the standard deviation is easily calculated (with the assumption of normality and by averaging the left and right distance to the mean). 
The resulting standard deviation of the distribution of $b_C$ is 0.13. 

\paragraph{River system driver}
\citet{Graf2006} analyzes the downstream hydrologic effects of 36 large dams in American rivers. 
In his Table 8 he provides regional values of the dam-capacity/yield ratio and of the percentage reduction in maximum annual discharge. 
Given that it is a large-scale study, we assume that the results are general enough to be reasonably transferred to our study region.
We assume that this reduction is registered right downstream of the dam (i.e. the ratio $A_i$/$A_T$ in Eq.9 is equal to 1), therefore it equals $\Delta RI$ (before and after the dam construction). 
We divide the reduction in maximum annual discharge by the capacity/yield ratio, to obtain regional estimates of the parameter $b_R$, and we consider the value corresponding to "all regions" (resulting equal to -0.30) as the mean of the prior distribution of $b_R$. 
We calculate the standard deviation of the $b_R$ values over the six regions in \citet{Graf2006} in order to obtain the standard deviation of the prior distribution of $b_R$ (resulting equal to 0.18).

The mean and standard deviation of the prior distribution of the parameters $b_A$, $b_C$ and $b_R$ are summarized in the third column of Table \ref{tab3}, with prior distribution assumed to be normal. 
Additional prior information is included about the shape of the prior distribution, based on the authors' understanding of the way the drivers may affect the magnitude of flood peaks.

\begin{figure}
\center\includegraphics[width=32pc]{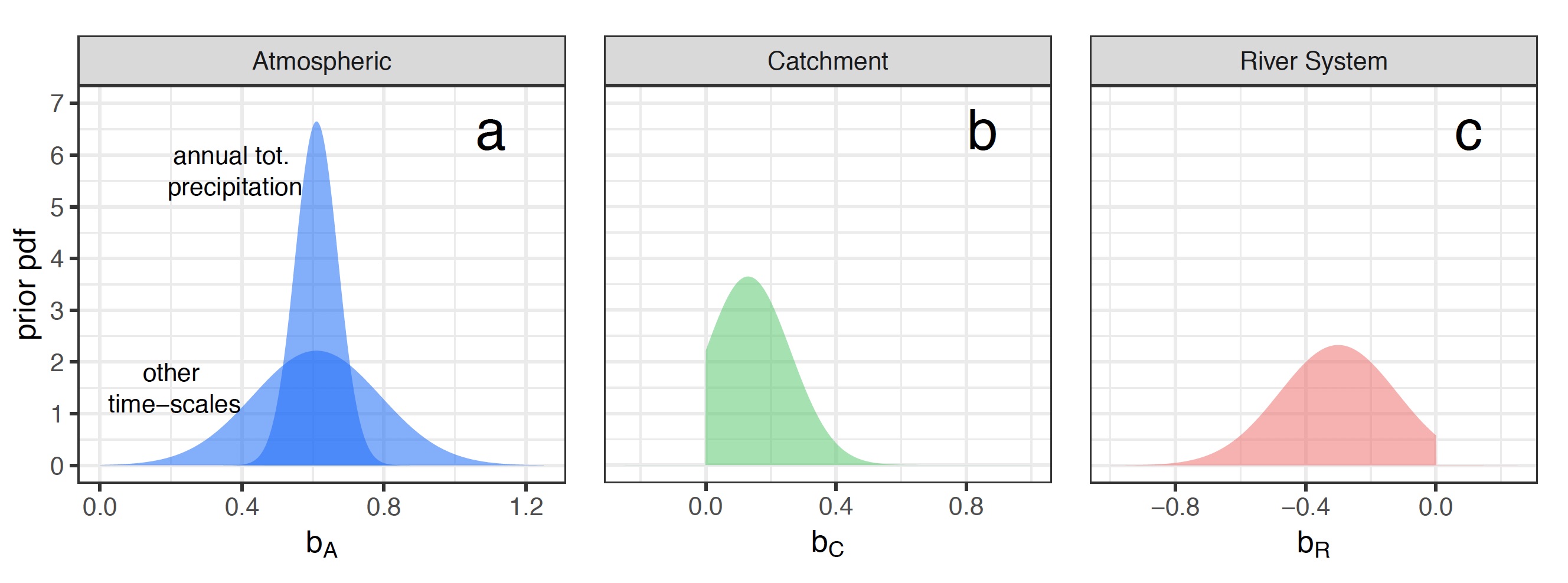}
\caption{Prior distribution of the model parameters $b_A$, $b_C$ and $b_R$, linking the changes of the drivers (i.e. the covariates of the alternative driver-informed models) to the changes of flood peaks. Each panel refers to a different driver (i.e. to a different driver-informed model): atmospheric driver (panel a), catchment driver (panel b) and river system driver (panel c). For the atmospheric driver we adopt different prior distributions for annual and extreme precipitation.}
\label{fig3}
\end{figure}

Increased (decreased) magnitude of flood peaks may result from an increase (a decrease) in the magnitude of precipitation. 
This is associated with a positive value of the regression parameter $b_A$ (i.e. the changes in the magnitude of flood peaks and in the covariate occur in the same direction/with the same sign). 
For this reason the lower tail of the prior normal distribution (contained in the third column of Table \ref{tab3}) of the parameter $b_A$ is truncated for negative values, in order to constrain the sign of the parameter. 
Similarly, we truncate the prior distribution of $b_C$ for negative values since soil degradation processes occurring in the catchment, associated with the intensification of agricultural practices, are expected to produce increased flooding. 
The construction of reservoirs (reflected in a positive step change in the reservoir index) may instead mitigate flood peaks in the downstream catchment. 
In this case the value of the parameter is negative and the upper tail of its prior normal distribution is truncated for positive values. 
The final types (lower- or upper- truncated normal) of the prior distribution of the regression parameters $b_A$, $b_C$ and $b_R$ are summarized in the fourth column of Table \ref{tab3} and represented in Figure \ref{fig3}.

\begin{table}
\small
\centering
\begin{tabular}{>{\raggedright\arraybackslash}p{2cm}>{\raggedright\arraybackslash}p{2.5cm}>{\raggedright\arraybackslash}p{3.5cm}>{\raggedright\arraybackslash}p{3.5cm}}
\hline
  Model and parameter & Study & Normal prior moments & Prior type\\
\hline
  $G_A$, $b_A$ & \citet{Perdigao2014}  & N(0.61, 0.06) with annual precipitation. N(0.61, 0.18) otherwise &  Truncated normal with lower tail truncated in 0\\ 
  $G_C$, $b_C$ & \citet{Fraser2013}    & N(0.13, 0.13) &  Truncated normal with lower tail truncated in 0\\
  $G_R$, $b_R$ & \citet{Graf2006}      & N(-0.30, 0.18) &  Truncated normal with upper tail truncated in 0\\
\hline
\end{tabular}
\caption{Sources, moments and type of the prior distribution of the model parameters $b_A$, $b_C$ and $b_R$.}
\label{tab3}
\end{table}

\section{Results}
In order to illustrate the methodology, we apply it first to one site (Section \ref{s:1catchment}).
The results for all other sites in Upper Austria are then presented in Section \ref{s:96catchments}.

\subsection{Attribution of flood changes in a single catchment} \label{s:1catchment}
\begin{figure}
\center\includegraphics[width=28pc]{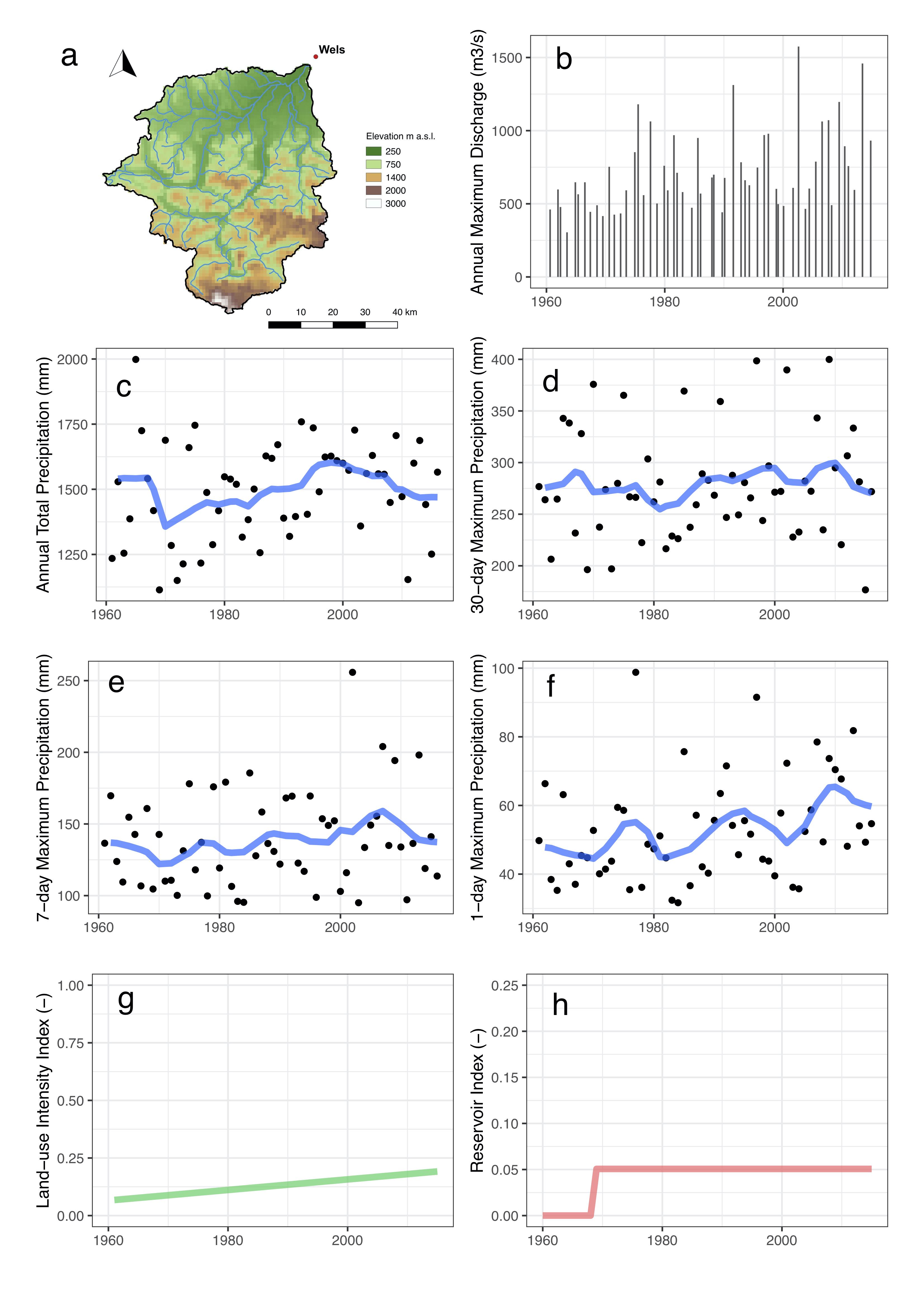}
\caption{\scriptsize{River Traun catchment, gauge station in Wels-Lichtenegg (panel a) and related flood series (panel b) and covariates representative for the three drivers of change: annual total precipitation (c), 30-day (d), 7-day (e) and 1-day maximum precipitation averaged over the catchment (f), land-use intensity index (g) and reservoir index (h).}}
\label{fig4}
\end{figure}

We analyze the river Traun catchment (gauge station in Wels-Lichtenegg, shown in panel a of Figure \ref{fig4}), where the AM series of flood peaks (panel b) presents a significant upward trend ($1.0 \pm 0.6 \%$ change per year). 
We apply the attribution framework in order to try to understand whether the magnitude of flood peaks is related to the temporal evolution of precipitation at the different time-scales (panels c, d, e and f), of the land-use intensity (panel g) or of the reservoir index (panel h) (i.e. if it can be attributed to one of the three drivers of change).
In particular, we assume that, the use of a covariate is informative if the WAIC value associated with the driver-informed model is lower than the one associated with the time-invariant model and their absolute difference is larger than a threshold, that we set to 2 using the same interpretation done with the AIC by \citet[pp.~700--71]{Burnham2002}.

Table \ref{tab4} shows the values of the WAIC associated with the alternative driver-informed models $G_A$, $G_C$, $G_R$ and the time-invariant $G_0$ in two cases: (i) when no prior information on the parameter $b$ is used (through a non-informative improper uniform distribution with infinite range), and (ii) with the priors of Figure \ref{fig3}. 
In the first case, by comparing the alternative models in terms of differences of WAIC (Table \ref{tab4}, first row), it emerges that the 1-day extreme precipitation (model $G_A$) and land-use intensity (model $G_R$) are the best covariates and the correspondent models outperform all others, including the time invariant model $G_0$.
This is because, as for the flood peak series, both 1-day extreme precipitation and  land-use intensity index have a positive trend over time (panels f and g).
Also the model $G_R$, that uses the reservoir index as covariate, provides a relatively good fit to the data (e.g. better than the time invariant model) since the Gmunden dam was built along the River Traun in 1969 (the location of the dam is shown in panel a of Figure \ref{fig4}), which is reflected in a step change in the reservoir index time series in the corresponding year (panel h). 

When prior information is used, the WAIC values (Table \ref{tab4}, second row) suggest that the model $G_A$ with the 1-day extreme precipitation is still the best one, but the models $G_C$ and $G_R$, using the land-use intensity and reservoir indexes, do not rank as well as they did before. 
This is because, in one case, crops cover less than 20\% of the total catchment area and, therefore, the land-use intensity varies in a low-value range. 
Crop areas are, in fact, concentrated in the northern part of the catchment, while the southern and middle part are mountainous areas (panel a of Figure \ref{fig4}).
In the other case, the reservoir index value after the dam construction ($\sim$0.05) is still significantly lower than the threshold value (0.25) between low and high flow alteration set by \citet{Lopez2013}. 
This is due to a small dam-capacity/mean-annual-flow-volume ratio.
In fact, the reservoir storage capacity (514$\times 10^6$ m$^3$) is significantly smaller than the mean annual flow volume of the catchment (4137$\times 10^6$ m$^3$), as well as the dam drainage area (1395 km$^2$) compared to the catchment area (3426 km$^2$).
Furthermore both flood peaks and the $RI$ increase in time, suggesting a positive value of the parameter $b_R$, which is in contrast with its informative prior distribution.

When using prior information on the parameter $b$ (see Figure \ref{fig3}), it becomes improbable that small values of the two indexes can produce significant flood changes, even though they vary in time in the same direction as the floods do (as in the case of the land-use intensity).
In this case, therefore, we attribute the temporal variability of floods to the long-term variation of the 1-day maximum precipitation. 

\begin{table}
\footnotesize
\centering
\begin{tabular}{>{\raggedright\arraybackslash}p{2cm}|>{\centering\arraybackslash}p{1.2cm}|>{\centering\arraybackslash}p{1.2cm}>{\centering\arraybackslash}p{1.2cm}>{\centering\arraybackslash}p{1.2cm}>{\centering\arraybackslash}p{1.2cm}|c|c}
\hline
 & $G_0$ & \multicolumn{4}{c|}{$G_A$} & $G_C$ & $G_R$ \\
 & Time-invariant & Annual Total P & 30-day maximum P & 7-day maximum P & 1-day maximum P & LI & RI \\ \hline
Non-informative priors & -126.9 & -125.0 & -125.2 & -127.7 & -133.4 & -133.0 & -130.0 \\ 
Informative priors &  & -126.6 & -127.1 & -129.1 & -133.7 & -127.6 & -126.2 \\ \hline
\end{tabular}
\caption{Comparison of the alternative time-invariant and driver-informed models for the river Traun catchment, gauge station in Wels-Lichtenegg. The values of the Widely-applicable information criterion, associated with each alternative model, are shown. The first row refers to the use of non-informative priors, while the second one refers to the priors of Table \ref{tab3}}

\label{tab4}
\end{table}

\subsection{Attribution of flood changes in Upper Austria} \label{s:96catchments}


In each of the 96 sites in Upper Austria the model $G_A$ is locally compared to the time-invariant model in terms of WAIC, which represents a trade-off between goodness of fit and model complexity.
We alternatively consider different time scales of precipitation as covariate of the driver-informed model.
In particular, we are interested in determining the most suitable time-scale for the atmospheric driver to be employed in the attribution study over the entire region, i.e. whether the long-term changes in annual precipitation or in the extreme precipitation drive flood changes in the region.

\begin{figure}
\center\includegraphics[width=32pc]{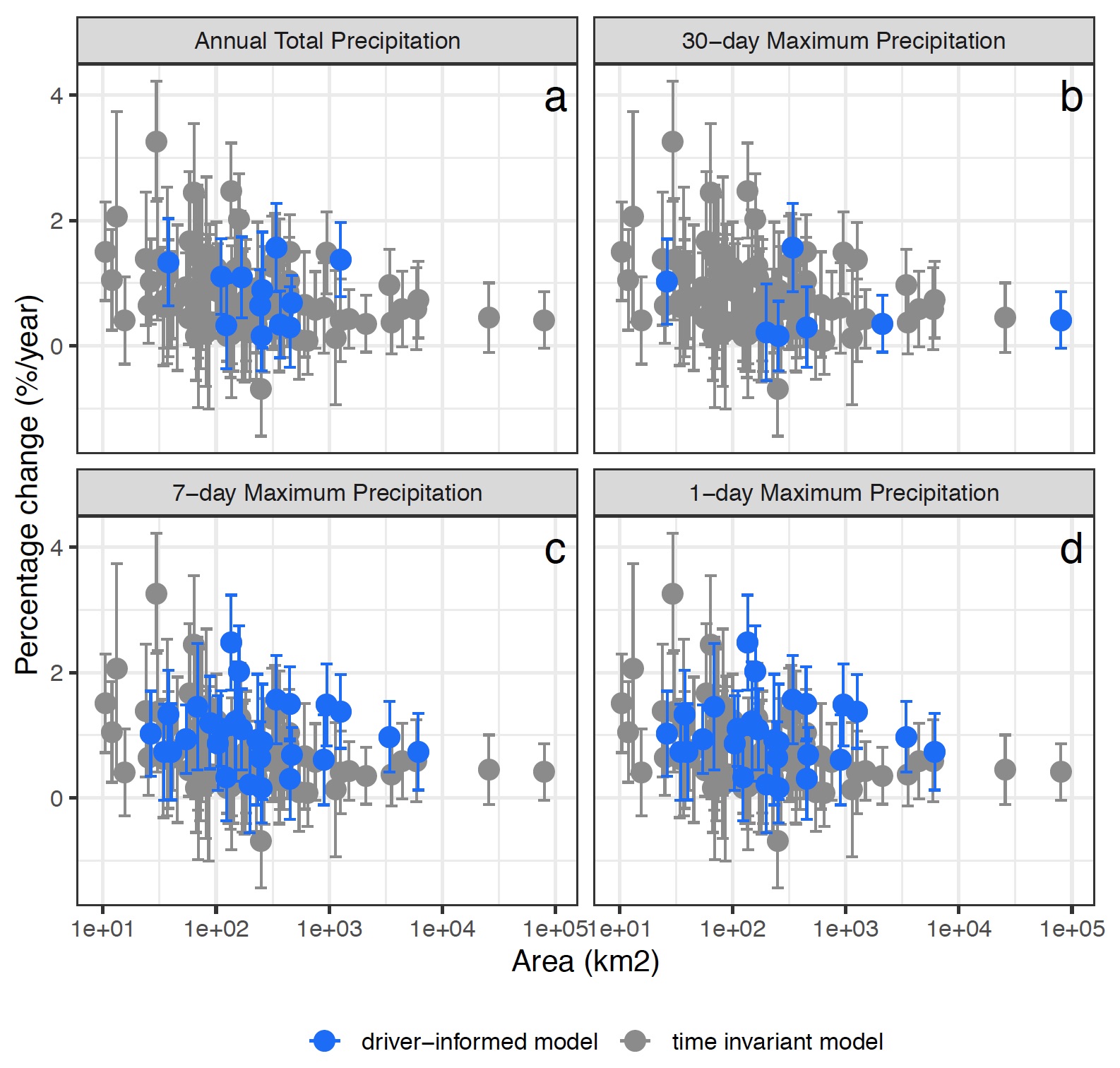}
\caption{Comparison between the driver-informed model (in blue), with precipitation as covariate, and the time-invariant model (in grey). 
The panels show the detected trends in flood series as a function of catchment area, with colors referring to the resulting best alternative model (i.e. time-invariant or driver-informed).
The selection of the best fitting model is carried out, in each site, through the Widely-Applicable information criterion. 
Each panel refers to a different time scale of precipitation used as covariate (annual total precipitation in panel a, 30-day maximum precipitation in panel b, 7-day maximum precipitation in panel c and 1-day maximum precipitation in panel d).}
\label{fig5}
\end{figure}

The results of this analysis are shown in Figure \ref{fig5} where, in each panel, a different time scale of the atmospheric driver is taken as covariate of the model $G_A$.
We mark the catchments in blue if the goodness of fit of the driver-informed model significantly improves with the inclusion of the covariate (accounting for the increased model complexity), with respect to the time-invariant case  (i.e. if $WAIC_{G_A}$ is lower than $WAIC_{G_0}$ and their absolute difference is larger than a threshold, arbitrarily set to 2).
Otherwise, we mark them in grey (meaning that the time-invariant model is still preferable).

The analysis shows that annual total precipitation as covariate  improves the model performance only for a small number of catchments in the region (panel a). On the contrary, extreme precipitation series with short durations (i.e. 7-day and 1-day maximum precipitation) seem to be regionally more suitable covariates for the distribution of AM (panels c and d). 

\begin{figure}
\center\includegraphics[width=32pc]{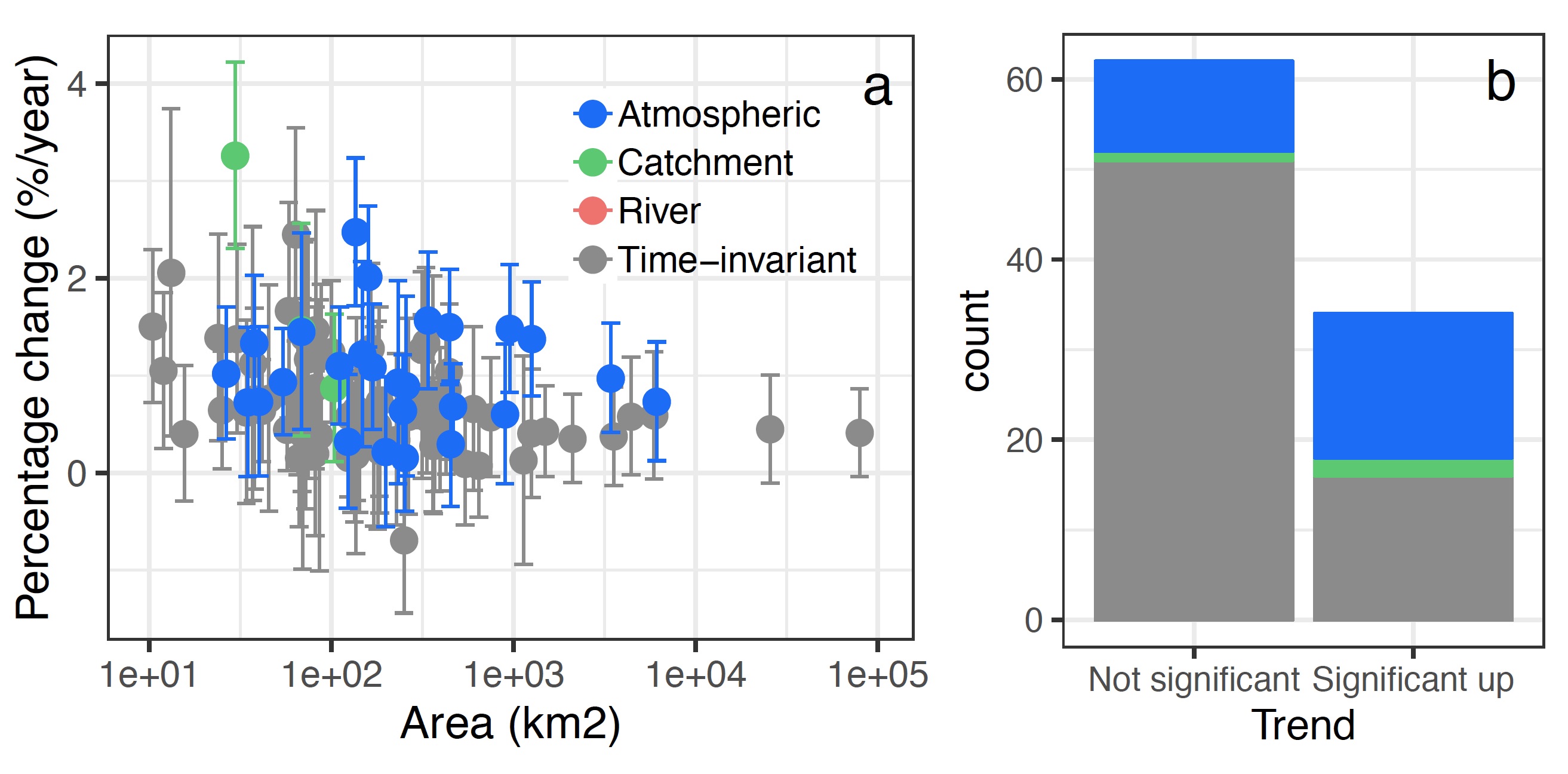}
\caption{Attribution of flood changes in Upper Austria to the atmospheric (blue), catchment (green) and river system driver (red).
Panel a shows the detected trends in flood series as a function of catchment area, with colors referring to the resulting best alternative driver-informed model. 
Catchments where the time-invariant model is still preferred are shown in grey. 
Panel b shows the occurrence of the selected alternative (driver-informed and time-invariant) models with a distinction between the catchments where the trends in flood peaks resulted significant (upward) or not significant.
The atmospheric driver is here represented by 1-day maximum precipitation.}
\label{fig6}
\end{figure}

Based on this analysis, we select 1-day maximum precipitation as covariate representative for the atmospheric processes driving flood change for the study region.
In each catchment we compare the WAIC values associated with four alternative models: $G_0$ (i.e. the time-invariant model), $G_A$ with 1-day maximum precipitation as covariate, $G_C$ and $G_R$.
Similarly to Figure \ref{fig5}, in Figure \ref{fig6} a catchment is marked in grey if the model $G_0$ is associated with the lowest value of WAIC.
Flood changes are instead attributed to one of the drivers (in Figure \ref{fig6} with colors) if the WAIC value of the corresponding driver-informed model is significantly lower than the one of the model $G_0$ (we use the same arbitrary threshold of WAIC difference equal to 2) and if it is the lowest among the competing driver-informed models. 

In a significant fraction of the catchments, the time-invariant model (in grey) is still the preferred choice while the atmospheric driver (in blue, represented by 1-day max precipitation as covariate) is the main driving process among the alternatives considered.
The catchment driver (in green) instead plays a very marginal role, together with the river system driver, which never results as best fitting model.
The long-term evolution of floods is attributed to the land-use intensification index only in three catchments with small catchment area (panel a).

\begin{figure}
\includegraphics[width=35pc]{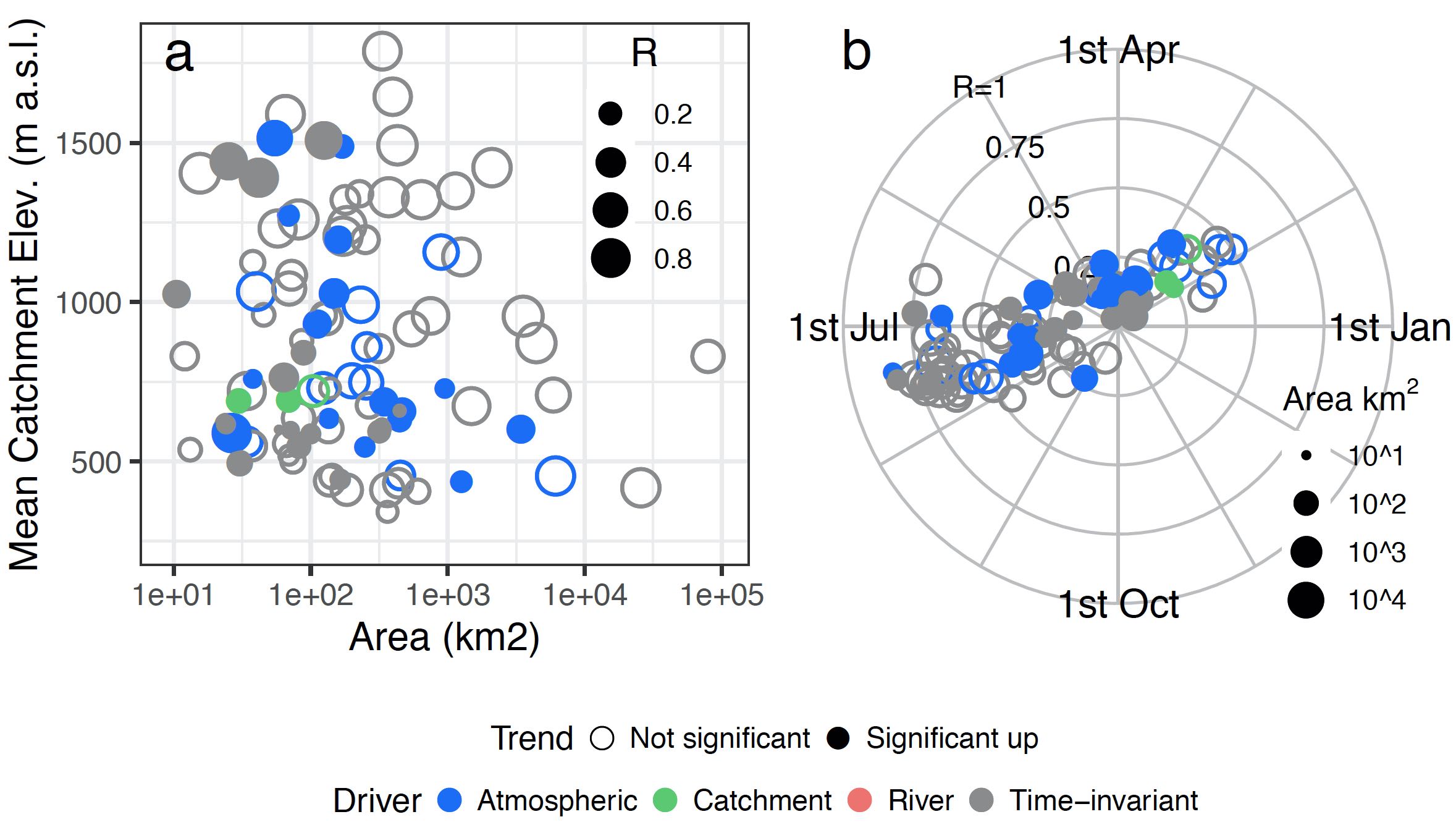} 
\caption{Mean catchment elevation as a function of catchment area (panel a) and seasonality of floods (panel b) in Upper Austria. The results of the attribution analysis (see Figure \ref{fig6}) are represented with colors and filled (empty) dots represent catchments with significant (not significant) flood trends. The size of the dots scales with the concentration of the date of occurrence of floods in panel a and with catchment area in panel b. The angular coordinate in panel b represents the average date of occurrence of floods and the distance from the center is the concentration of the date of occurrence $R$ ($R=0$ when floods are evenly distributed throughout the year and $R=1$ when all floods occur on the same day). Both are calculated as in \citet[]{Bloschl2017}.}
\label{fig7}
\end{figure}

Panel b shows the occurrence of the attributed drivers with a distinction between the catchments where the trends in time of flood peaks resulted significant or not significant (see Figure \ref{fig2}). 
The flood series in around half of the sites, where trends in time of the floods are significant, are associated to the long-term evolution of extreme precipitation series.
However, the other half of them does not correlate significantly with any of the covariates used here, even though the correlation with time is significant. 
All of these sites have relatively small catchments and one third of them are in the mountains (Figure \ref{fig7}a). Figure \ref{fig7}b shows that, in terms of seasonality of floods, the sites with trends but no correlated covariate are not significantly different from the others.

\begin{figure}
\center\includegraphics[width=32pc]{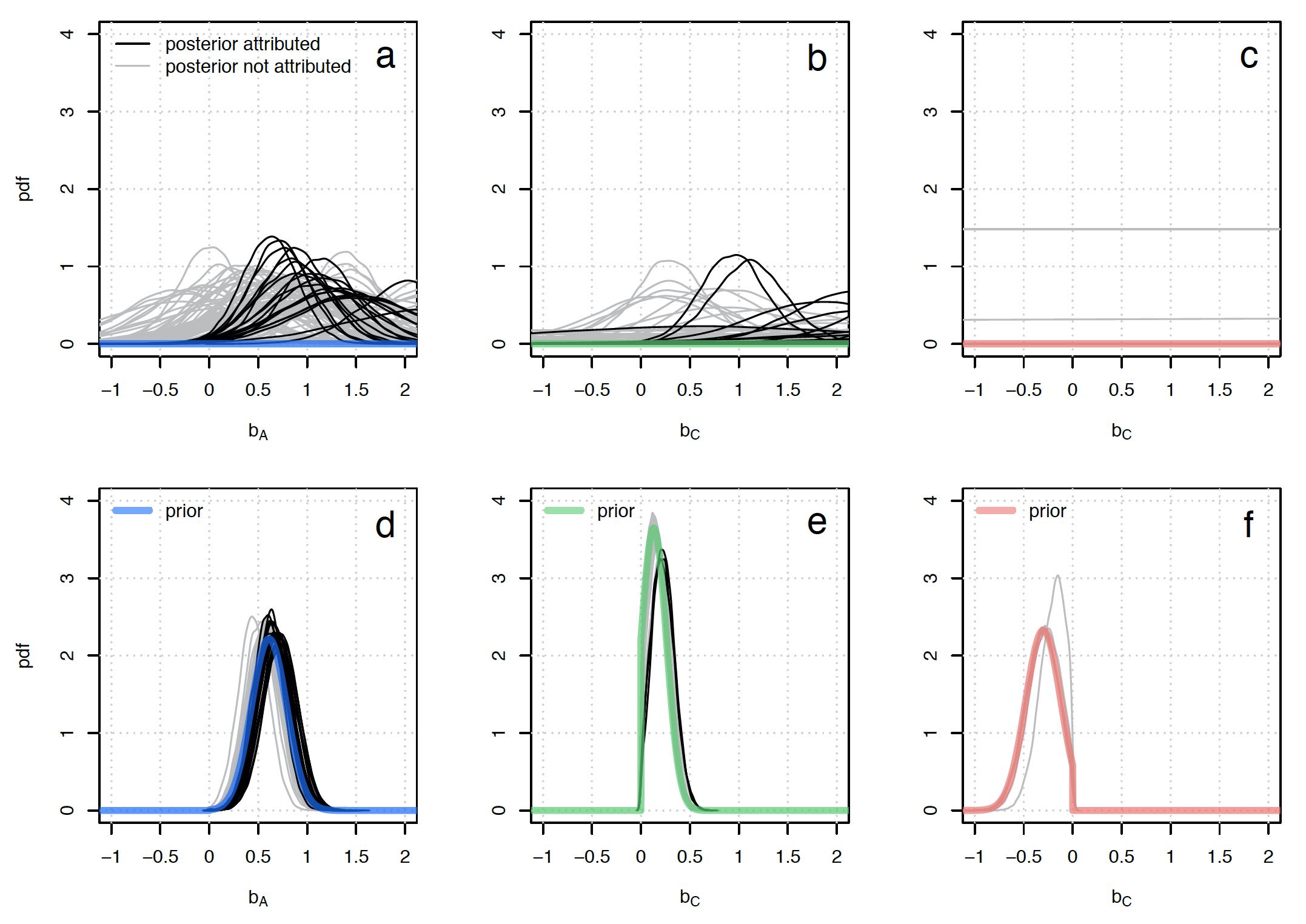}
\caption{Prior distribution of the regression parameters $b_A$ (Atmospheric driver, panels a and d), $b_C$ (Catchment driver, panel b and e) and $b_R$ (River system driver, panel c and f) with the corresponding posterior distributions for each catchment. Upper panels refer to the use of non-informative priors and lower panels of the informative priors of Figure \ref{fig3}. When the evolution of flood peaks in one catchment is attributed to one driver, the posterior distribution of the corresponding parameter is shown in black, otherwise in grey.}
\label{fig8}
\end{figure}

Figure \ref{fig8} compares the posterior distribution of the parameters $b_A$, $b_C$ and $b_R$, obtained with the MCMC approach, to their corresponding prior distribution. 
When the evolution of flood peaks in one catchment is attributed to one driver, the posterior distribution of the corresponding regression parameter is represented in black, otherwise (i.e. if the flood changes are attributed to other drivers or the time-invariant model is preferred) in grey.
In the upper panels non-informative priors are used while, in the lower panels, the informative priors, shown in Figure \ref{fig3}, are used, consistently with Figure \ref{fig5} and \ref{fig6}.
This figure shows the influence of the informative priors in the attribution process. 
By introducing additional external information about how the connection between these covariates and flood peaks should be, we obtain very different posterior estimates of the parameters $b$ and, consequently, of the extreme value distribution parameters and of the attribution results.

\begin{figure}
\center\includegraphics[width=32pc]{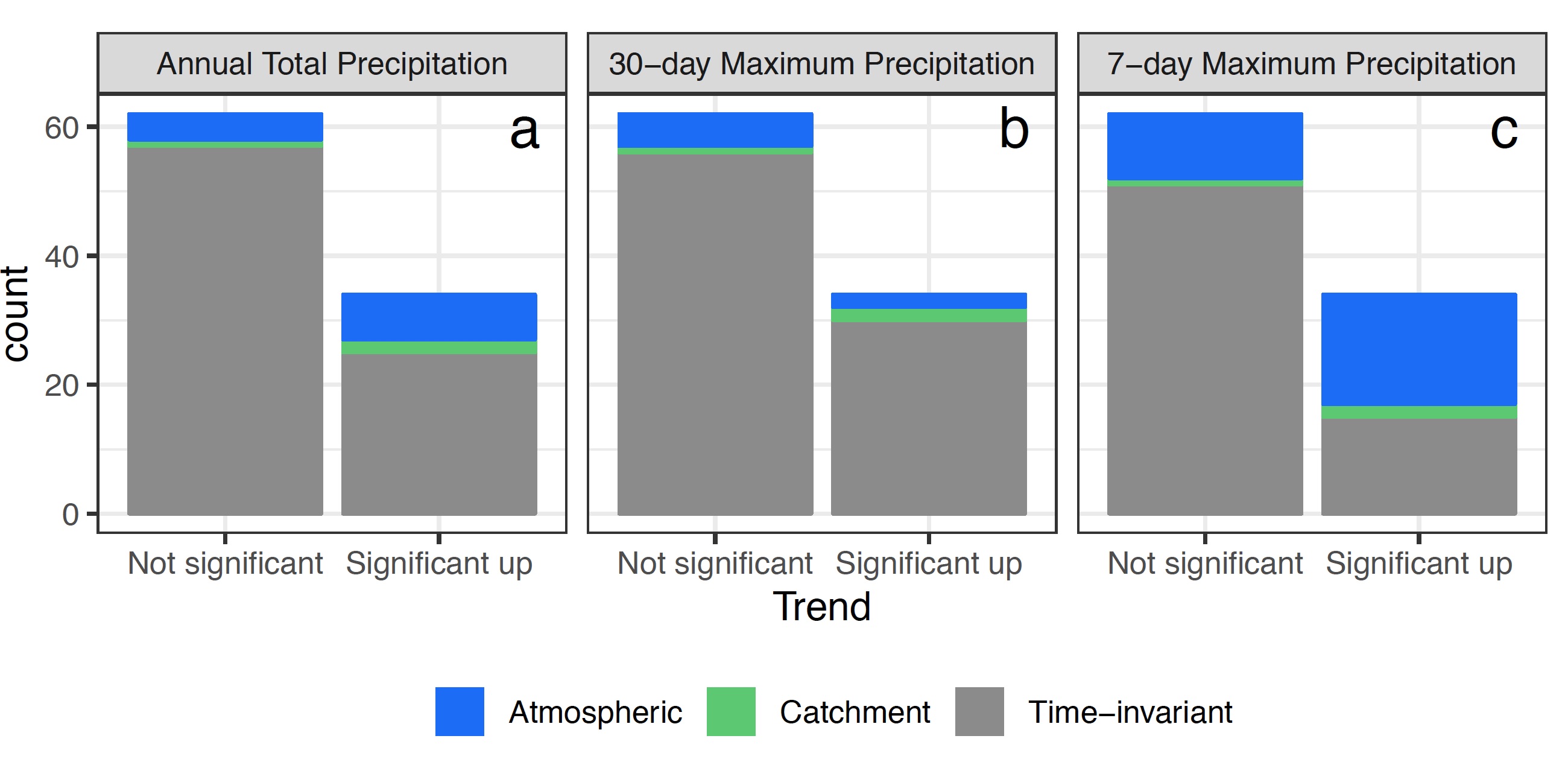}
\caption{Same as panel b of Figure \ref{fig6} but for different time scales of precipitation. Occurrence of the selected alternative (driver-informed and time-invariant) models is shown, with a distinction between the catchments where the trends in flood peaks resulted significant (upward) or not significant. The considered precipitation time-scales for the atmospheric driver are: annual precipitation (panel a), 30-day maximum precipitation (panel b) and 7-day maximum precipitation (panel c).}
\label{fig9}
\end{figure}

Similarly to panel b of Figure \ref{fig6}, Figure \ref{fig9} shows the number of occurrence of attributed driver types for the other precipitation time-scales. 
Different covariates (annual precipitation, 30-day maximum precipitation and 7-day maximum precipitation) for the model $G_A$ are considered in the different panels. 
The changes in the decadal annual precipitation correspond to only around one fourth of the significant trends in time detected in flood series (even less for the 30-day maximum precipitation). 
The 7-day maximum precipitation series as covariate show instead a similar results as the 1-day maximum precipitation (see figure \ref{fig6}, panel b).

\section{Discussion and conclusions}

In this study we apply a simple data-based approach for the attribution of flood changes to potential drivers: atmospheric, catchment and river system drivers. 
The method is applied to a large number of catchments in a study region, Upper Austria, where significant positive trends are detected in maximum annual peak discharge series. 
We assume the maximum annual peak discharges to follow a two-parameter Gumbel distribution. 
We include information on the three drivers through covariates (smoothed/decadal annual precipitation, smoothed/decadal 30-day, 7-day, 1-day maximum annual precipitation, land-use index and reservoir index) that control the location parameter of the Gumbel distribution through simple log-linear and log-log models.
The attribution is performed by comparing the different models, using different covariates, fitted using Bayesian inference. 
The comparison is based on the trade-off between goodness of fit and model complexity, using the Watanabe-Akaike information criterion (WAIC).
Prior information on the slope parameters of these models (i.e. on the elasticity of the covariates to floods), based on results of published studies, is also provided in order to limit the possibility for spurious correlations to bias the attribution. 
Without using information on the expected elasticity, the attribution procedure is ill posed in that it would prefer the covariate better correlated to the flood temporal fluctuations, no matter if the correlation is physically plausible.

Our results suggest that precipitation change is the main driver of flood change in the study region (no matter which time-scale is used for precipitation), which is consistent with the results in \citet{Viglione2016}.
Differently from what suggested in \citet{Sivapalan2015} and \citet{Sraj2016}, annual precipitation is not as good as extreme precipitation in explaining the long-term evolution of floods in this context. 
This is due to the fact that, while \citet{Sraj2016} are interested in how floods correlate to precipitation at the annual scale, here we are looking at long-term (decadal) variation of precipitation.
The smoothing of the annual precipitation time series results in averaging wet years and dry years, thus destroying the correlation to floods.
On the contrary, the extreme precipitation series, even after the smoothing, do not contain the influence of droughts and are therefore more correlated to long-term fluctuations of the flood statistics.
In Upper Austria, because of the relatively small size of the catchments, the 7-day and the 1-day maximum annual precipitation decadal fluctuations correlate best with the fluctuations of the flood statistics.

Land-use intensity changes are significant in very few small catchments, which are mostly covered by agricultural land.
Differently from what has been assumed in \citet{Viglione2016}, these are not the smallest catchments, which are located in the mountains where there is almost no agriculture and there has not been a significant deforestation nor afforestation in the last 50 years.
For most of the catchments, land-use intensity changes (note that we investigated the changes related to late-harvested crops, see Section 3.2) do not correlate meaningfully with flood changes (we get a good correlation only if we use non-informative priors for the elasticity parameter, resulting in not credible posterior distributions). 
This is consistent with the fact that, in Upper Austria, big floods occur generally in summer, in correspondence of precipitation events with high magnitude, and smaller floods are in spring or winter.
Few floods occur in autumn, when we would expect a greater soil susceptibility to erosion and compaction (potentially leading to increased flooding) as a consequence of the agricultural practices for late-harvested crops \citep{Chamen2003, Beven2008}.

Reservoirs do not produce relevant effects on floods neither, because the capacity/yield ratio is generally small. 
Most of the dams are built for hydroelectricity purposes, but even for those built for flood control we do not detect significant flood attenuation at the gauging stations because these effects are mainly local \citep{Ayalew2017, Volpi2018}.
This result is not surprising given that we expect reservoirs to attenuate flood peaks and that we observe mostly upward trends in flood peak magnitude in the region.

In half of the catchments where we detect significant trends in flood peaks, the driver-informed model, with extreme precipitation as covariate, outperforms the time-invariant model. 
In the other cases we observe significant trends but not a significant correlation to the covariates, suggesting that the long-term temporal evolution of the selected drivers is overall not sufficient to explain the observed trends in the peak discharge series and that other covariates should be considered or covariates informative on other drivers of flood change.
For example, we did not consider changes in snow related processes here (e.g. by taking air temperature as covariate), which may be important for mountainous catchments \citep[see e.g.][]{Bloschl2017}, and changes in precipitation of shorter durations (e.g. hourly precipitation), which may be more appropriate covariate for the smaller catchments. 
Indeed, all of the sites where we do detect a trend in flood peaks but no correlation with the covariates are small (and some mountainous) catchments.
The fact that in these catchments we have not identified a suitable driver may also suggest that other flood-driver relations should be explored in future analyses, representing for example the combined effect of multiple drivers on flood change.

In some of the catchments where we do not detect significant trends in flood peaks, the driver-informed model, with extreme precipitation as covariate, outperforms the time-invariant model. Through the driver informed models used here, long term flood fluctuations are related to the covariates, even in cases where no monotonic trend in time is detected.
This is in line with our objective to research the relationships between flood temporal variations and the long-term evolution of the drivers.

This study considers many sites in one region, but the analysis is essentially local, i.e. every site is analysed independently using locally defined covariates.
There is potential for extending the method to something in line with \citet{Viglione2016}, in which a regional model is fitted to all the sites jointly explicitly using covariates for the drivers.

The framework used here is easily generalizable and applicable in other contexts (i.e. by changing the covariates or the model structure).
Different drivers could be considered, that may have positive or negative effects on floods.
The key issue, as shown in this paper, is to gather prior information on how sensitive are floods to changes in the drivers, which could be achieved through derived-distribution \citep[see e.g.][]{Eagleson1972WRR, Sivapalanetal2005WRR, Volpi2018} and comparative process studies \citep[see e.g.][]{FalkenmarkChapman1989, Viglione2013, PUBbook2013}. 
This is in line with the concept of Flood Frequency Hydrology \citep{MerzBloeschl2008aWRR, MerzBloeschl2008bWRR, Viglioneetal2013WRR}, which highlights the importance of combining flood data with additional types of information, including causal mechanisms, to improve flood frequency estimation and, as in this case, to support change analyses.

\section*{Acknowledgments}
This project has received funding from the European Union’s Horizon 2020 research and innovation programme under the Marie Sk\l{}odowska-Curie grant agreement No 676027 and from the Austrian Science Funds (project I 3174).

This product incorporates data from the GRanD database which is \textcopyright Global Water System Project (2011).



\bibliographystyle{elsarticle-harv} 
\bibliography{attribution}

\begin{thebibliography}{67}
\expandafter\ifx\csname natexlab\endcsname\relax\def\natexlab#1{#1}\fi
\providecommand{\url}[1]{\texttt{#1}}
\providecommand{\href}[2]{#2}
\providecommand{\path}[1]{#1}
\providecommand{\DOIprefix}{doi:}
\providecommand{\ArXivprefix}{arXiv:}
\providecommand{\URLprefix}{URL: }
\providecommand{\Pubmedprefix}{pmid:}
\providecommand{\doi}[1]{\href{http://dx.doi.org/#1}{\path{#1}}}
\providecommand{\Pubmed}[1]{\href{pmid:#1}{\path{#1}}}
\providecommand{\bibinfo}[2]{#2}
\ifx\xfnm\relax \def\xfnm[#1]{\unskip,\space#1}\fi
\bibitem[{Akaike(1973)}]{akaike1973information}
\bibinfo{author}{Akaike, H.}, \bibinfo{year}{1973}.
\newblock \bibinfo{title}{Information theory and an extension of the maximum
  likelihood principle}, in: \bibinfo{booktitle}{Selected Papers of Hirotugu
  Akaike}. \bibinfo{publisher}{Springer New York}, \bibinfo{address}{New York,
  NY}, pp. \bibinfo{pages}{199--213}.
\bibitem[{Alaoui et~al.(2018)Alaoui, Rogger, Peth and
  Bl{\"{o}}schl}]{Alaoui2018}
\bibinfo{author}{Alaoui, A.}, \bibinfo{author}{Rogger, M.},
  \bibinfo{author}{Peth, S.}, \bibinfo{author}{Bl{\"{o}}schl, G.},
  \bibinfo{year}{2018}.
\newblock \bibinfo{title}{{Does soil compaction increase floods? A review}}.
\newblock \bibinfo{journal}{Journal of Hydrology} \bibinfo{volume}{557},
  \bibinfo{pages}{631--642}.
\newblock \DOIprefix\doi{10.1016/j.jhydrol.2017.12.052}.
\bibitem[{Archfield et~al.(2016)Archfield, Hirsch, Viglione and
  Bl{\"{o}}schl}]{Archfield2016}
\bibinfo{author}{Archfield, S.A.}, \bibinfo{author}{Hirsch, R.M.},
  \bibinfo{author}{Viglione, A.}, \bibinfo{author}{Bl{\"{o}}schl, G.},
  \bibinfo{year}{2016}.
\newblock \bibinfo{title}{{Fragmented patterns of flood change across the
  United States}}.
\newblock \bibinfo{journal}{Geophysical Research Letters} \bibinfo{volume}{43},
  \bibinfo{pages}{10,232--10,239}.
\newblock \DOIprefix\doi{10.1002/2016GL070590}.
\bibitem[{Ayalew et~al.(2017)Ayalew, Krajewski, Mantilla, Wright and
  Small}]{Ayalew2017}
\bibinfo{author}{Ayalew, T.B.}, \bibinfo{author}{Krajewski, W.F.},
  \bibinfo{author}{Mantilla, R.}, \bibinfo{author}{Wright, D.B.},
  \bibinfo{author}{Small, S.J.}, \bibinfo{year}{2017}.
\newblock \bibinfo{title}{{Effect of Spatially Distributed Small Dams on Flood
  Frequency: Insights from the Soap Creek Watershed}}.
\newblock \bibinfo{journal}{Journal of Hydrologic Engineering}
  \bibinfo{volume}{22}, \bibinfo{pages}{04017011}.
\newblock \DOIprefix\doi{10.1061/(ASCE)HE.1943-5584.0001513}.
\bibitem[{Beven et~al.(2008)Beven, Young, Romanowicz, O'Connell, Ewen,
  O'Donnell, Homan, Posthumus, Morris, Hollis, Rose, Lamb and
  Archer}]{Beven2008}
\bibinfo{author}{Beven, K.J.}, \bibinfo{author}{Young, P.C.},
  \bibinfo{author}{Romanowicz, R.J.}, \bibinfo{author}{O'Connell, E.},
  \bibinfo{author}{Ewen, J.}, \bibinfo{author}{O'Donnell, G.},
  \bibinfo{author}{Homan, I.}, \bibinfo{author}{Posthumus, H.},
  \bibinfo{author}{Morris, J.}, \bibinfo{author}{Hollis, J.},
  \bibinfo{author}{Rose, S.}, \bibinfo{author}{Lamb, R.},
  \bibinfo{author}{Archer, D.}, \bibinfo{year}{2008}.
\newblock \bibinfo{title}{{FD2120: Analysis of historical data sets to look for
  impacts of land use management change on flood generation}}.
\newblock \bibinfo{type}{Technical Report}. Defra/EA.
\bibitem[{Bl{\"{o}}schl et~al.(2007)Bl{\"{o}}schl, Ardoin-Bardin, Bonell,
  Dorninger, Goodrich, Gutknecht, Matamoros, Merz, Shand and
  Szolgay}]{Bloschl2007}
\bibinfo{author}{Bl{\"{o}}schl, G.}, \bibinfo{author}{Ardoin-Bardin, S.},
  \bibinfo{author}{Bonell, M.}, \bibinfo{author}{Dorninger, M.},
  \bibinfo{author}{Goodrich, D.}, \bibinfo{author}{Gutknecht, D.},
  \bibinfo{author}{Matamoros, D.}, \bibinfo{author}{Merz, B.},
  \bibinfo{author}{Shand, P.}, \bibinfo{author}{Szolgay, J.},
  \bibinfo{year}{2007}.
\newblock \bibinfo{title}{{At what scales do climate variability and land cover
  change impact on flooding and low flows?}}
\newblock \bibinfo{journal}{Hydrological Processes} \bibinfo{volume}{21},
  \bibinfo{pages}{1241--1247}.
\newblock \DOIprefix\doi{10.1002/hyp.6669}.
\bibitem[{Bl{\"{o}}schl et~al.(2017)Bl{\"{o}}schl, Hall, Parajka,
  Perdig{\~{a}}o, Merz, Arheimer, Aronica, Bilibashi, Bonacci, Borga,
  {\v{C}}anjevac, Castellarin, Chirico, Claps, Fiala, Frolova, Gorbachova,
  G{\"{u}}l, Hannaford, Harrigan, Kireeva, Kiss, Kjeldsen, Kohnov{\'{a}},
  Koskela, Ledvinka, Macdonald, Mavrova-Guirguinova, Mediero, Merz, Molnar,
  Montanari, Murphy, Osuch, Ovcharuk, Radevski, Rogger, Salinas, Sauquet,
  {\v{S}}raj, Szolgay, Viglione, Volpi, Wilson, Zaimi and
  {\v{Z}}ivkovi{\'{c}}}]{Bloschl2017}
\bibinfo{author}{Bl{\"{o}}schl, G.}, \bibinfo{author}{Hall, J.},
  \bibinfo{author}{Parajka, J.}, \bibinfo{author}{Perdig{\~{a}}o, R.A.P.},
  \bibinfo{author}{Merz, B.}, \bibinfo{author}{Arheimer, B.},
  \bibinfo{author}{Aronica, G.T.}, \bibinfo{author}{Bilibashi, A.},
  \bibinfo{author}{Bonacci, O.}, \bibinfo{author}{Borga, M.},
  \bibinfo{author}{{\v{C}}anjevac, I.}, \bibinfo{author}{Castellarin, A.},
  \bibinfo{author}{Chirico, G.B.}, \bibinfo{author}{Claps, P.},
  \bibinfo{author}{Fiala, K.}, \bibinfo{author}{Frolova, N.},
  \bibinfo{author}{Gorbachova, L.}, \bibinfo{author}{G{\"{u}}l, A.},
  \bibinfo{author}{Hannaford, J.}, \bibinfo{author}{Harrigan, S.},
  \bibinfo{author}{Kireeva, M.}, \bibinfo{author}{Kiss, A.},
  \bibinfo{author}{Kjeldsen, T.R.}, \bibinfo{author}{Kohnov{\'{a}}, S.},
  \bibinfo{author}{Koskela, J.J.}, \bibinfo{author}{Ledvinka, O.},
  \bibinfo{author}{Macdonald, N.}, \bibinfo{author}{Mavrova-Guirguinova, M.},
  \bibinfo{author}{Mediero, L.}, \bibinfo{author}{Merz, R.},
  \bibinfo{author}{Molnar, P.}, \bibinfo{author}{Montanari, A.},
  \bibinfo{author}{Murphy, C.}, \bibinfo{author}{Osuch, M.},
  \bibinfo{author}{Ovcharuk, V.}, \bibinfo{author}{Radevski, I.},
  \bibinfo{author}{Rogger, M.}, \bibinfo{author}{Salinas, J.L.},
  \bibinfo{author}{Sauquet, E.}, \bibinfo{author}{{\v{S}}raj, M.},
  \bibinfo{author}{Szolgay, J.}, \bibinfo{author}{Viglione, A.},
  \bibinfo{author}{Volpi, E.}, \bibinfo{author}{Wilson, D.},
  \bibinfo{author}{Zaimi, K.}, \bibinfo{author}{{\v{Z}}ivkovi{\'{c}}, N.},
  \bibinfo{year}{2017}.
\newblock \bibinfo{title}{{Changing climate shifts timing of European floods}}.
\newblock \bibinfo{journal}{Science} \bibinfo{volume}{357},
  \bibinfo{pages}{588--590}.
\newblock \DOIprefix\doi{10.1126/science.aan2506}.
\bibitem[{Bl{\"o}schl et~al.(2012)Bl{\"o}schl, Merz, Parajka, Salinas and
  Viglione}]{Bloschl2012}
\bibinfo{author}{Bl{\"o}schl, G.}, \bibinfo{author}{Merz, R.},
  \bibinfo{author}{Parajka, J.}, \bibinfo{author}{Salinas, J.L.},
  \bibinfo{author}{Viglione, A.}, \bibinfo{year}{2012}.
\newblock \bibinfo{title}{Floods in austria}.
\newblock \bibinfo{journal}{IAHS Spec. Publ.} \bibinfo{volume}{10},
  \bibinfo{pages}{169--177}.
\bibitem[{Bl{\"o}schl et~al.(2011)Bl{\"o}schl, Viglione, Merz, Parajka, Salinas
  and Sch{\"o}ner}]{Bloschl2011}
\bibinfo{author}{Bl{\"o}schl, G.}, \bibinfo{author}{Viglione, A.},
  \bibinfo{author}{Merz, R.}, \bibinfo{author}{Parajka, J.},
  \bibinfo{author}{Salinas, J.L.}, \bibinfo{author}{Sch{\"o}ner, W.},
  \bibinfo{year}{2011}.
\newblock \bibinfo{title}{{Auswirkungen des Klimawandels auf Hochwasser und
  Niederwasser}}.
\newblock \bibinfo{journal}{{\"O}sterreichische Wasser- und Abfallwirtschaft}
  \bibinfo{volume}{63}, \bibinfo{pages}{21--30}.
\newblock \DOIprefix\doi{10.1007/s00506-010-0269-z}.
\bibitem[{Blöschl et~al.(2013)Blöschl, Sivapalan, Wagener, Viglione and
  Savenije}]{PUBbook2013}
\bibinfo{author}{Blöschl, G.}, \bibinfo{author}{Sivapalan, M.},
  \bibinfo{author}{Wagener, T.}, \bibinfo{author}{Viglione, A.},
  \bibinfo{author}{Savenije, H.H.}, \bibinfo{year}{2013}.
\newblock \bibinfo{title}{{Runoff Prediction in Ungauged Basins - Synthesis
  across Processes, Places and Scales}}.
\newblock \bibinfo{publisher}{Cambridge University Press}.
\newblock \URLprefix \url{http://www.cambridge.org/9781107028180}.
\bibitem[{Bronstert et~al.(2007)Bronstert, B{\'{a}}rdossy, Bismuth, Buiteveld,
  Disse, Engel, Fritsch, Hundecha, Lammersen, Niehoff and
  Ritter}]{Bronstert2007}
\bibinfo{author}{Bronstert, A.}, \bibinfo{author}{B{\'{a}}rdossy, A.},
  \bibinfo{author}{Bismuth, C.}, \bibinfo{author}{Buiteveld, H.},
  \bibinfo{author}{Disse, M.}, \bibinfo{author}{Engel, H.},
  \bibinfo{author}{Fritsch, U.}, \bibinfo{author}{Hundecha, Y.},
  \bibinfo{author}{Lammersen, R.}, \bibinfo{author}{Niehoff, D.},
  \bibinfo{author}{Ritter, N.}, \bibinfo{year}{2007}.
\newblock \bibinfo{title}{{Multi-scale modelling of land-use change and river
  training effects on floods in the Rhine basin}}.
\newblock \bibinfo{journal}{River Research and Applications}
  \bibinfo{volume}{23}, \bibinfo{pages}{1102--1125}.
\newblock \DOIprefix\doi{10.1002/rra}.
\bibitem[{Burnham and Anderson(2002)}]{Burnham2002}
\bibinfo{author}{Burnham, K.}, \bibinfo{author}{Anderson, D.},
  \bibinfo{year}{2002}.
\newblock \bibinfo{title}{{Model Selection and Multimodel Inference: A
  Practical Information-Theoretic Approach (2nd ed)}}. volume
  \bibinfo{volume}{172}.
\newblock \URLprefix
  \url{http://linkinghub.elsevier.com/retrieve/pii/S0304380003004526},
  \DOIprefix\doi{10.1016/j.ecolmodel.2003.11.004},
  \href{http://arxiv.org/abs/arXiv:1011.1669v3}{{\tt arXiv:arXiv:1011.1669v3}}.
\bibitem[{Carpenter et~al.(2017)Carpenter, Gelman, Hoffman, Lee, Goodrich,
  Betancourt, Brubaker, Guo, Li and Riddell}]{Carpenter2017}
\bibinfo{author}{Carpenter, B.}, \bibinfo{author}{Gelman, A.},
  \bibinfo{author}{Hoffman, M.}, \bibinfo{author}{Lee, D.},
  \bibinfo{author}{Goodrich, B.}, \bibinfo{author}{Betancourt, M.},
  \bibinfo{author}{Brubaker, M.}, \bibinfo{author}{Guo, J.},
  \bibinfo{author}{Li, P.}, \bibinfo{author}{Riddell, A.},
  \bibinfo{year}{2017}.
\newblock \bibinfo{title}{Stan: A probabilistic programming language}.
\newblock \bibinfo{journal}{Journal of Statistical Software, Articles}
  \bibinfo{volume}{76}, \bibinfo{pages}{1--32}.
\newblock \URLprefix \url{https://www.jstatsoft.org/v076/i01},
  \DOIprefix\doi{10.18637/jss.v076.i01}.
\bibitem[{Chamen et~al.(2003)Chamen, Alakukku, Pires, Sommer, Spoor, Tijink and
  Weisskopf}]{Chamen2003}
\bibinfo{author}{Chamen, T.}, \bibinfo{author}{Alakukku, L.},
  \bibinfo{author}{Pires, S.}, \bibinfo{author}{Sommer, C.},
  \bibinfo{author}{Spoor, G.}, \bibinfo{author}{Tijink, F.},
  \bibinfo{author}{Weisskopf, P.}, \bibinfo{year}{2003}.
\newblock \bibinfo{title}{Prevention strategies for field traffic-induced
  subsoil compaction : a review ; part 2, equipment and field practices}.
\newblock \bibinfo{journal}{Soil \& tillage research : an international journal
  on research and development in soil tillage and field traffic, and their
  relationship with land use, crop production and the environment}
  \bibinfo{volume}{73}, \bibinfo{pages}{161--174}.
\bibitem[{Cleveland(1979)}]{Cleveland1979}
\bibinfo{author}{Cleveland, W.S.}, \bibinfo{year}{1979}.
\newblock \bibinfo{title}{Robust locally weighted regression and smoothing
  scatterplots}.
\newblock \bibinfo{journal}{Journal of the American Statistical Association}
  \bibinfo{volume}{74}, \bibinfo{pages}{829--836}.
\newblock \DOIprefix\doi{10.1080/01621459.1979.10481038}.
\bibitem[{Dietrich et~al.(2012)Dietrich, Schmitz, M{\"{u}}ller, Fader,
  Lotze-Campen and Popp}]{Dietrich2012}
\bibinfo{author}{Dietrich, J.P.}, \bibinfo{author}{Schmitz, C.},
  \bibinfo{author}{M{\"{u}}ller, C.}, \bibinfo{author}{Fader, M.},
  \bibinfo{author}{Lotze-Campen, H.}, \bibinfo{author}{Popp, A.},
  \bibinfo{year}{2012}.
\newblock \bibinfo{title}{{Measuring agricultural land-use intensity - A global
  analysis using a model-assisted approach}}.
\newblock \bibinfo{journal}{Ecological Modelling} \bibinfo{volume}{232},
  \bibinfo{pages}{109--118}.
\newblock \DOIprefix\doi{10.1016/j.ecolmodel.2012.03.002}.
\bibitem[{Eagleson(1972)}]{Eagleson1972WRR}
\bibinfo{author}{Eagleson, P.S.}, \bibinfo{year}{1972}.
\newblock \bibinfo{title}{{Dynamics of Flood Frequency}}.
\newblock \bibinfo{journal}{Water Resources Research} \bibinfo{volume}{8},
  \bibinfo{pages}{878--\&}.
\bibitem[{Falkenmark and Chapman(1989)}]{FalkenmarkChapman1989}
\bibinfo{author}{Falkenmark, M.}, \bibinfo{author}{Chapman, T.},
  \bibinfo{year}{1989}.
\newblock \bibinfo{title}{{Comparative hydrology: An ecological approach to
  land and water resources}}.
\newblock \bibinfo{publisher}{The Unesco Press}, \bibinfo{address}{Paris}.
\bibitem[{Fraser et~al.(2013)Fraser, Mcintyre, Jackson and
  Wheater}]{Fraser2013}
\bibinfo{author}{Fraser, C.E.}, \bibinfo{author}{Mcintyre, N.},
  \bibinfo{author}{Jackson, B.M.}, \bibinfo{author}{Wheater, H.S.},
  \bibinfo{year}{2013}.
\newblock \bibinfo{title}{{Upscaling hydrological processes and land management
  change impacts using a metamodeling procedure}}.
\newblock \bibinfo{journal}{Water Resources Research} \bibinfo{volume}{49},
  \bibinfo{pages}{5817--5833}.
\newblock \DOIprefix\doi{10.1002/wrcr.20432}.
\bibitem[{Gelman et~al.(2014)Gelman, Hwang and Vehtari}]{Gelman2014}
\bibinfo{author}{Gelman, A.}, \bibinfo{author}{Hwang, J.},
  \bibinfo{author}{Vehtari, A.}, \bibinfo{year}{2014}.
\newblock \bibinfo{title}{{Understanding predictive information criteria for
  Bayesian models}}.
\newblock \bibinfo{journal}{Statistics and Computing} \bibinfo{volume}{24},
  \bibinfo{pages}{997--1016}.
\newblock \DOIprefix\doi{10.1007/s11222-013-9416-2},
  \href{http://arxiv.org/abs/1307.5928}{{\tt arXiv:1307.5928}}.
\bibitem[{Graf(2006)}]{Graf2006}
\bibinfo{author}{Graf, W.L.}, \bibinfo{year}{2006}.
\newblock \bibinfo{title}{{Downstream hydrologic and geomorphic effects of
  large dams on American rivers}}.
\newblock \bibinfo{journal}{Geomorphology} \bibinfo{volume}{79},
  \bibinfo{pages}{336--360}.
\newblock \DOIprefix\doi{10.1016/j.geomorph.2006.06.022}.
\bibitem[{Hall et~al.(2014)Hall, Arheimer, Borga, Br{\'{a}}zdil, Claps, Kiss,
  Kjeldsen, Kriau{\v{c}}iūnienė, Kundzewicz, Lang, Llasat, Macdonald,
  McIntyre, Mediero, Merz, Merz, Molnar, Montanari, Neuhold, Parajka,
  Perdig{\~{a}}o, Plavcov{\'{a}}, Rogger, Salinas, Sauquet, Sch{\"{a}}r,
  Szolgay, Viglione and Bl{\"{o}}schl}]{Hall2014}
\bibinfo{author}{Hall, J.}, \bibinfo{author}{Arheimer, B.},
  \bibinfo{author}{Borga, M.}, \bibinfo{author}{Br{\'{a}}zdil, R.},
  \bibinfo{author}{Claps, P.}, \bibinfo{author}{Kiss, A.},
  \bibinfo{author}{Kjeldsen, T.R.}, \bibinfo{author}{Kriau{\v{c}}iūnienė,
  J.}, \bibinfo{author}{Kundzewicz, Z.W.}, \bibinfo{author}{Lang, M.},
  \bibinfo{author}{Llasat, M.C.}, \bibinfo{author}{Macdonald, N.},
  \bibinfo{author}{McIntyre, N.}, \bibinfo{author}{Mediero, L.},
  \bibinfo{author}{Merz, B.}, \bibinfo{author}{Merz, R.},
  \bibinfo{author}{Molnar, P.}, \bibinfo{author}{Montanari, A.},
  \bibinfo{author}{Neuhold, C.}, \bibinfo{author}{Parajka, J.},
  \bibinfo{author}{Perdig{\~{a}}o, R.A.P.}, \bibinfo{author}{Plavcov{\'{a}},
  L.}, \bibinfo{author}{Rogger, M.}, \bibinfo{author}{Salinas, J.L.},
  \bibinfo{author}{Sauquet, E.}, \bibinfo{author}{Sch{\"{a}}r, C.},
  \bibinfo{author}{Szolgay, J.}, \bibinfo{author}{Viglione, A.},
  \bibinfo{author}{Bl{\"{o}}schl, G.}, \bibinfo{year}{2014}.
\newblock \bibinfo{title}{{Understanding flood regime changes in Europe: a
  state-of-the-art assessment}}.
\newblock \bibinfo{journal}{Hydrology and Earth System Sciences}
  \bibinfo{volume}{18}, \bibinfo{pages}{2735--2772}.
\newblock \DOIprefix\doi{10.5194/hess-18-2735-2014}.
\bibitem[{Hiebl and Frei(2018)}]{Hiebl2018}
\bibinfo{author}{Hiebl, J.}, \bibinfo{author}{Frei, C.}, \bibinfo{year}{2018}.
\newblock \bibinfo{title}{Daily precipitation grids for austria since
  1961---development and evaluation of a spatial dataset for hydroclimatic
  monitoring and modelling}.
\newblock \bibinfo{journal}{Theoretical and Applied Climatology}
  \bibinfo{volume}{132}, \bibinfo{pages}{327--345}.
\newblock \DOIprefix\doi{10.1007/s00704-017-2093-x}.
\bibitem[{Krumphuber(2016)}]{lko2016}
\bibinfo{author}{Krumphuber, C.}, \bibinfo{year}{2016}.
\newblock \bibinfo{title}{{Crop farming in Upper Austria}}.
\newblock \bibinfo{type}{Technical Report}. Landwirtschaftskammer
  Ober{\"{o}}sterreich.
\newblock \bibinfo{note}{\url{www.ooe.lko.at}}.
\bibitem[{Lammersen et~al.(2002)Lammersen, Engel, van~de Langemheen and
  Buiteveld}]{Lammersen2002}
\bibinfo{author}{Lammersen, R.}, \bibinfo{author}{Engel, H.},
  \bibinfo{author}{van~de Langemheen, W.}, \bibinfo{author}{Buiteveld, H.},
  \bibinfo{year}{2002}.
\newblock \bibinfo{title}{{Impact of river training and retention measures on
  flood peaks along the Rhine}}.
\newblock \bibinfo{journal}{Journal of Hydrology} \bibinfo{volume}{267},
  \bibinfo{pages}{115--124}.
\newblock \DOIprefix\doi{10.1016/S0022-1694(02)00144-0}.
\bibitem[{Lehner et~al.(2011)Lehner, Liermann, Revenga, V{\"{o}}r{\"{o}}smarty,
  Fekete, Crouzet, D{\"{o}}ll, Endejan, Frenken, Magome, Nilsson, Robertson,
  R{\"{o}}del, Sindorf and Wisser}]{grand2011}
\bibinfo{author}{Lehner, B.}, \bibinfo{author}{Liermann, C.R.},
  \bibinfo{author}{Revenga, C.}, \bibinfo{author}{V{\"{o}}r{\"{o}}smarty, C.},
  \bibinfo{author}{Fekete, B.}, \bibinfo{author}{Crouzet, P.},
  \bibinfo{author}{D{\"{o}}ll, P.}, \bibinfo{author}{Endejan, M.},
  \bibinfo{author}{Frenken, K.}, \bibinfo{author}{Magome, J.},
  \bibinfo{author}{Nilsson, C.}, \bibinfo{author}{Robertson, J.C.},
  \bibinfo{author}{R{\"{o}}del, R.}, \bibinfo{author}{Sindorf, N.},
  \bibinfo{author}{Wisser, D.}, \bibinfo{year}{2011}.
\newblock \bibinfo{title}{{High-resolution mapping of the world's reservoirs
  and dams for sustainable river-flow management}}.
\newblock \bibinfo{journal}{Frontiers in Ecology and the Environment}
  \bibinfo{volume}{9}, \bibinfo{pages}{494--502}.
\newblock \DOIprefix\doi{10.1890/100125}.
\bibitem[{L{\'{o}}pez and Franc{\'{e}}s(2013)}]{Lopez2013}
\bibinfo{author}{L{\'{o}}pez, J.}, \bibinfo{author}{Franc{\'{e}}s, F.},
  \bibinfo{year}{2013}.
\newblock \bibinfo{title}{{Non-stationary flood frequency analysis in
  continental Spanish rivers, using climate and reservoir indices as external
  covariates}}.
\newblock \bibinfo{journal}{Hydrology and Earth System Sciences}
  \bibinfo{volume}{17}, \bibinfo{pages}{3189--3203}.
\newblock \DOIprefix\doi{10.5194/hess-17-3189-2013}.
\bibitem[{Mangini et~al.(2018)Mangini, Viglione, Hall, Hundecha, Ceola,
  Montanari, Rogger, Salinas, Borz{\`{i}} and Parajka}]{Mangini2018}
\bibinfo{author}{Mangini, W.}, \bibinfo{author}{Viglione, A.},
  \bibinfo{author}{Hall, J.}, \bibinfo{author}{Hundecha, Y.},
  \bibinfo{author}{Ceola, S.}, \bibinfo{author}{Montanari, A.},
  \bibinfo{author}{Rogger, M.}, \bibinfo{author}{Salinas, J.L.},
  \bibinfo{author}{Borz{\`{i}}, I.}, \bibinfo{author}{Parajka, J.},
  \bibinfo{year}{2018}.
\newblock \bibinfo{title}{{Detection of trends in magnitude and frequency of
  flood peaks across Europe}}.
\newblock \bibinfo{journal}{Hydrological Sciences Journal}
  \bibinfo{volume}{63}, \bibinfo{pages}{1--20}.
\newblock \DOIprefix\doi{10.1080/02626667.2018.1444766}.
\bibitem[{Mediero et~al.(2014)Mediero, Santill{\'{a}}n, Garrote and
  Granados}]{Mediero2014}
\bibinfo{author}{Mediero, L.}, \bibinfo{author}{Santill{\'{a}}n, D.},
  \bibinfo{author}{Garrote, L.}, \bibinfo{author}{Granados, A.},
  \bibinfo{year}{2014}.
\newblock \bibinfo{title}{{Detection and attribution of trends in magnitude,
  frequency and timing of floods in Spain}}.
\newblock \bibinfo{journal}{Journal of Hydrology} \bibinfo{volume}{517},
  \bibinfo{pages}{1072--1088}.
\newblock \DOIprefix\doi{10.1016/j.jhydrol.2014.06.040}.
\bibitem[{Merz et~al.(2012)Merz, Vorogushyn, Uhlemann, Delgado and
  Hundecha}]{Merz2012}
\bibinfo{author}{Merz, B.}, \bibinfo{author}{Vorogushyn, S.},
  \bibinfo{author}{Uhlemann, S.}, \bibinfo{author}{Delgado, J.},
  \bibinfo{author}{Hundecha, Y.}, \bibinfo{year}{2012}.
\newblock \bibinfo{title}{{HESS Opinions "More efforts and scientific rigour
  are needed to attribute trends in flood time series"}}.
\newblock \bibinfo{journal}{Hydrology and Earth System Sciences}
  \bibinfo{volume}{16}, \bibinfo{pages}{1379--1387}.
\newblock \DOIprefix\doi{10.5194/hess-16-1379-2012}.
\bibitem[{Merz and Bl{\"o}schl(2008a)}]{MerzBloeschl2008aWRR}
\bibinfo{author}{Merz, R.}, \bibinfo{author}{Bl{\"o}schl, G.},
  \bibinfo{year}{2008}a.
\newblock \bibinfo{title}{{Flood frequency hydrology: 1. Temporal, spatial, and
  causal expansion of information}}.
\newblock \bibinfo{journal}{Water Resources Research} \bibinfo{volume}{44}.
\newblock \DOIprefix\doi{10.1029/2007WR006744}.
\bibitem[{Merz and Bl{\"o}schl(2008b)}]{MerzBloeschl2008bWRR}
\bibinfo{author}{Merz, R.}, \bibinfo{author}{Bl{\"o}schl, G.},
  \bibinfo{year}{2008}b.
\newblock \bibinfo{title}{{Flood frequency hydrology: 2. Combining data
  evidence}}.
\newblock \bibinfo{journal}{Water Resources Research} \bibinfo{volume}{44}.
\newblock \DOIprefix\doi{10.1029/2007WR006745}.
\bibitem[{Monfreda et~al.(2008)Monfreda, Ramankutty and Foley}]{Monfreda2008}
\bibinfo{author}{Monfreda, C.}, \bibinfo{author}{Ramankutty, N.},
  \bibinfo{author}{Foley, J.A.}, \bibinfo{year}{2008}.
\newblock \bibinfo{title}{{Farming the planet: 2. Geographic distribution of
  crop areas, yields, physiological types, and net primary production in the
  year 2000}}.
\newblock \bibinfo{journal}{Global Biogeochemical Cycles} \bibinfo{volume}{22},
  \bibinfo{pages}{1--19}.
\newblock \DOIprefix\doi{10.1029/2007GB002947}.
\bibitem[{Montanari and Koutsoyiannis(2014)}]{Montanari2014}
\bibinfo{author}{Montanari, A.}, \bibinfo{author}{Koutsoyiannis, D.},
  \bibinfo{year}{2014}.
\newblock \bibinfo{title}{{Modeling and mitigating natural hazards:
  Stationarity is immortal!}}
\newblock \bibinfo{journal}{Water Resources Research} \bibinfo{volume}{50},
  \bibinfo{pages}{9748--9756}.
\newblock \DOIprefix\doi{10.1002/2014WR016092}.
\bibitem[{Mudelsee et~al.(2003)Mudelsee, B{\"{o}}rngen, Tetzlaff and
  Gr{\"{u}}newald}]{Mudelsee2003}
\bibinfo{author}{Mudelsee, M.}, \bibinfo{author}{B{\"{o}}rngen, M.},
  \bibinfo{author}{Tetzlaff, G.}, \bibinfo{author}{Gr{\"{u}}newald, U.},
  \bibinfo{year}{2003}.
\newblock \bibinfo{title}{{No upward trends in the occurrence of extreme floods
  in central Europe}}.
\newblock \bibinfo{journal}{Nature} \bibinfo{volume}{425},
  \bibinfo{pages}{166--169}.
\newblock \DOIprefix\doi{10.1038/nature01928}.
\bibitem[{Niehoff et~al.(2002)Niehoff, Fritsch and Bronstert}]{Niehoff2002}
\bibinfo{author}{Niehoff, D.}, \bibinfo{author}{Fritsch, U.},
  \bibinfo{author}{Bronstert, A.}, \bibinfo{year}{2002}.
\newblock \bibinfo{title}{{Land-use impacts on storm-runoff generation:
  Scenarios of land-use change and simulation of hydrological response in a
  meso-scale catchment in SW-Germany}}.
\newblock \bibinfo{journal}{Journal of Hydrology} \bibinfo{volume}{267},
  \bibinfo{pages}{80--93}.
\newblock \DOIprefix\doi{10.1016/S0022-1694(02)00142-7}.
\bibitem[{O'Connell et~al.(2007)O'Connell, Ewen, O'Donnell and
  Quinn}]{OConnell2007}
\bibinfo{author}{O'Connell, P.E.}, \bibinfo{author}{Ewen, J.},
  \bibinfo{author}{O'Donnell, G.}, \bibinfo{author}{Quinn, P.},
  \bibinfo{year}{2007}.
\newblock \bibinfo{title}{{Is there a link between agricultural land-use
  management and flooding?}}
\newblock \bibinfo{journal}{Hydrology and Earth System Sciences}
  \bibinfo{volume}{11}, \bibinfo{pages}{96--107}.
\newblock \DOIprefix\doi{10.5194/hess-11-96-2007}.
\bibitem[{Perdig{\~{a}}o and Bl{\"{o}}schl(2014)}]{Perdigao2014}
\bibinfo{author}{Perdig{\~{a}}o, R.A.}, \bibinfo{author}{Bl{\"{o}}schl, G.},
  \bibinfo{year}{2014}.
\newblock \bibinfo{title}{{Spatiotemporal flood sensitivity to annual
  precipitation: Evidence for landscape-climate coevolution}}.
\newblock \bibinfo{journal}{Water Resources Research} \bibinfo{volume}{50},
  \bibinfo{pages}{5492--5509}.
\newblock \DOIprefix\doi{10.1002/2014WR015365}.
\bibitem[{Petrow and Merz(2009)}]{Petrow2009}
\bibinfo{author}{Petrow, T.}, \bibinfo{author}{Merz, B.}, \bibinfo{year}{2009}.
\newblock \bibinfo{title}{{Trends in flood magnitude, frequency and seasonality
  in Germany in the period 1951–2002}}.
\newblock \bibinfo{journal}{Journal of Hydrology} \bibinfo{volume}{371},
  \bibinfo{pages}{129--141}.
\newblock \DOIprefix\doi{10.1016/j.jhydrol.2009.03.024}.
\bibitem[{Pinter et~al.(2006)Pinter, {Van der Ploeg}, Schweigert and
  Hoefer}]{Pinter2006}
\bibinfo{author}{Pinter, N.}, \bibinfo{author}{{Van der Ploeg}, R.R.},
  \bibinfo{author}{Schweigert, P.}, \bibinfo{author}{Hoefer, G.},
  \bibinfo{year}{2006}.
\newblock \bibinfo{title}{{Flood magnification on the River Rhine}}.
\newblock \bibinfo{journal}{Hydrological Processes} \bibinfo{volume}{20},
  \bibinfo{pages}{147--164}.
\newblock \DOIprefix\doi{10.1002/hyp.5908}.
\bibitem[{van~der Ploeg et~al.(2002)van~der Ploeg, Machulla, Hermsmeyer,
  Ilsemann, Gieska and Bachmann}]{VanderPloeg2002}
\bibinfo{author}{van~der Ploeg, R.}, \bibinfo{author}{Machulla, G.},
  \bibinfo{author}{Hermsmeyer, D.}, \bibinfo{author}{Ilsemann, J.},
  \bibinfo{author}{Gieska, M.}, \bibinfo{author}{Bachmann, J.},
  \bibinfo{year}{2002}.
\newblock \bibinfo{title}{{Changes in land use and the growing number of flash
  floods in Germany}}.
\newblock \bibinfo{journal}{Agricultural Effects on Ground and Surface Waters:
  Research at the Edge of Science and Society} , \bibinfo{pages}{317--321}.
\bibitem[{Prosdocimi et~al.(2014)Prosdocimi, Kjeldsen and
  Svensson}]{Prosdocimi2014}
\bibinfo{author}{Prosdocimi, I.}, \bibinfo{author}{Kjeldsen, T.R.},
  \bibinfo{author}{Svensson, C.}, \bibinfo{year}{2014}.
\newblock \bibinfo{title}{{Non-stationarity in annual and seasonal series of
  peak flow and precipitation in the UK}}.
\newblock \bibinfo{journal}{Natural Hazards and Earth System Sciences}
  \bibinfo{volume}{14}, \bibinfo{pages}{1125--1144}.
\newblock \DOIprefix\doi{10.5194/nhess-14-1125-2014}.
\bibitem[{Ramankutty et~al.(2008)Ramankutty, Evan, Monfreda and
  Foley}]{Ramankutty2008}
\bibinfo{author}{Ramankutty, N.}, \bibinfo{author}{Evan, A.T.},
  \bibinfo{author}{Monfreda, C.}, \bibinfo{author}{Foley, J.A.},
  \bibinfo{year}{2008}.
\newblock \bibinfo{title}{{Farming the planet: 1. Geographic distribution of
  global agricultural lands in the year 2000}}.
\newblock \bibinfo{journal}{Global Biogeochemical Cycles} \bibinfo{volume}{22},
  \bibinfo{pages}{1--19}.
\newblock \DOIprefix\doi{10.1029/2007GB002952}.
\bibitem[{Ray et~al.(2012)Ray, Ramankutty, Mueller, West and Foley}]{Ray2012}
\bibinfo{author}{Ray, D.K.}, \bibinfo{author}{Ramankutty, N.},
  \bibinfo{author}{Mueller, N.D.}, \bibinfo{author}{West, P.C.},
  \bibinfo{author}{Foley, J.A.}, \bibinfo{year}{2012}.
\newblock \bibinfo{title}{{Recent patterns of crop yield growth and
  stagnation}}.
\newblock \bibinfo{journal}{Nature Communications} \bibinfo{volume}{3},
  \bibinfo{pages}{1293--1297}.
\newblock \DOIprefix\doi{10.1038/ncomms2296}.
\bibitem[{Renard and Lall(2014)}]{Renard2014}
\bibinfo{author}{Renard, B.}, \bibinfo{author}{Lall, U.}, \bibinfo{year}{2014}.
\newblock \bibinfo{title}{{Regional frequency analysis conditioned on
  large-scale atmospheric or oceanic fields}}.
\newblock \bibinfo{journal}{Water Resources Research} \bibinfo{volume}{50},
  \bibinfo{pages}{9536--9554}.
\newblock \DOIprefix\doi{10.1002/2014WR016277},
  \href{http://arxiv.org/abs/2014WR016527}{{\tt arXiv:2014WR016527}}.
\bibitem[{Rogger et~al.(2017)Rogger, Agnoletti, Alaoui, Bathurst, Bodner,
  Borga, Chaplot, Gallart, Glatzel, Hall, Holden, Holko, Horn, Kiss, Quinton,
  Leitinger, Lennartz, Parajka, Peth, Robinson, Salinas, Santoro, Szolgay, Tron
  and Viglione}]{Rogger2017}
\bibinfo{author}{Rogger, M.}, \bibinfo{author}{Agnoletti, M.},
  \bibinfo{author}{Alaoui, A.}, \bibinfo{author}{Bathurst, J.C.},
  \bibinfo{author}{Bodner, G.}, \bibinfo{author}{Borga, M.},
  \bibinfo{author}{Chaplot, V.}, \bibinfo{author}{Gallart, F.},
  \bibinfo{author}{Glatzel, G.}, \bibinfo{author}{Hall, J.},
  \bibinfo{author}{Holden, J.}, \bibinfo{author}{Holko, L.},
  \bibinfo{author}{Horn, R.}, \bibinfo{author}{Kiss, A.},
  \bibinfo{author}{Quinton, J.N.}, \bibinfo{author}{Leitinger, G.},
  \bibinfo{author}{Lennartz, B.}, \bibinfo{author}{Parajka, J.},
  \bibinfo{author}{Peth, S.}, \bibinfo{author}{Robinson, M.},
  \bibinfo{author}{Salinas, J.L.}, \bibinfo{author}{Santoro, A.},
  \bibinfo{author}{Szolgay, J.}, \bibinfo{author}{Tron, S.},
  \bibinfo{author}{Viglione, A.}, \bibinfo{year}{2017}.
\newblock \bibinfo{title}{{Land use change impacts on floods at the catchment
  scale: Challenges and opportunities for future research}}.
\newblock \bibinfo{journal}{Water Resouces Research} \bibinfo{volume}{53},
  \bibinfo{pages}{5209--5219}.
\newblock \DOIprefix\doi{10.1002/2017WR020723.Received}.
\bibitem[{Salazar et~al.(2012)Salazar, Frances, Komma, Blume, Francke,
  Bronstert and Bloschl}]{Salazar2012}
\bibinfo{author}{Salazar, S.}, \bibinfo{author}{Frances, F.},
  \bibinfo{author}{Komma, J.}, \bibinfo{author}{Blume, T.},
  \bibinfo{author}{Francke, T.}, \bibinfo{author}{Bronstert, A.},
  \bibinfo{author}{Bloschl, G.}, \bibinfo{year}{2012}.
\newblock \bibinfo{title}{{A comparative analysis of the effectiveness of flood
  management measures based on the concept of "retaining water in the
  landscape" in different European hydro-climatic regions}}.
\newblock \bibinfo{journal}{Natural Hazards and Earth System Science}
  \bibinfo{volume}{12}, \bibinfo{pages}{3287--3306}.
\newblock \DOIprefix\doi{10.5194/nhess-12-3287-2012}.
\bibitem[{Serinaldi and Kilsby(2015)}]{Serinaldi2015}
\bibinfo{author}{Serinaldi, F.}, \bibinfo{author}{Kilsby, C.G.},
  \bibinfo{year}{2015}.
\newblock \bibinfo{title}{{Stationarity is undead: Uncertainty dominates the
  distribution of extremes}}.
\newblock \bibinfo{journal}{Advances in Water Resources} \bibinfo{volume}{77},
  \bibinfo{pages}{17--36}.
\newblock \DOIprefix\doi{10.1016/j.advwatres.2014.12.013}.
\bibitem[{Silva et~al.(2017)Silva, Portela, Naghettini and
  Fernandes}]{Silva2017}
\bibinfo{author}{Silva, A.T.}, \bibinfo{author}{Portela, M.M.},
  \bibinfo{author}{Naghettini, M.}, \bibinfo{author}{Fernandes, W.},
  \bibinfo{year}{2017}.
\newblock \bibinfo{title}{{A Bayesian peaks-over-threshold analysis of floods
  in the Itaja{\'{i}}-a{\c{c}}u River under stationarity and nonstationarity}}.
\newblock \bibinfo{journal}{Stochastic Environmental Research and Risk
  Assessment} \bibinfo{volume}{31}, \bibinfo{pages}{185--204}.
\newblock \DOIprefix\doi{10.1007/s00477-015-1184-4}.
\bibitem[{Sivapalan and Bl{\"{o}}schl(2015)}]{Sivapalan2015}
\bibinfo{author}{Sivapalan, M.}, \bibinfo{author}{Bl{\"{o}}schl, G.},
  \bibinfo{year}{2015}.
\newblock \bibinfo{title}{{Time scale interactions and the coevolution of
  humans and water}}.
\newblock \bibinfo{journal}{Water Resources Research} \bibinfo{volume}{51},
  \bibinfo{pages}{6988--7022}.
\newblock \DOIprefix\doi{10.1002/2015WR017896},
  \href{http://arxiv.org/abs/2014WR016527}{{\tt arXiv:2014WR016527}}.
\bibitem[{Sivapalan et~al.(2005)Sivapalan, Bl{\"o}schl, Merz and
  Gutknecht}]{Sivapalanetal2005WRR}
\bibinfo{author}{Sivapalan, M.}, \bibinfo{author}{Bl{\"o}schl, G.},
  \bibinfo{author}{Merz, R.}, \bibinfo{author}{Gutknecht, D.},
  \bibinfo{year}{2005}.
\newblock \bibinfo{title}{{Linking flood frequency to long-term water balance:
  Incorporating effects of seasonality}}.
\newblock \bibinfo{journal}{Water Resources Research} \bibinfo{volume}{41}.
\newblock \DOIprefix\doi{10.1029/2004WR003439}.
\bibitem[{Skublics et~al.(2016)Skublics, Bl{\"{o}}schl and
  Rutschmann}]{Skublics2016}
\bibinfo{author}{Skublics, D.}, \bibinfo{author}{Bl{\"{o}}schl, G.},
  \bibinfo{author}{Rutschmann, P.}, \bibinfo{year}{2016}.
\newblock \bibinfo{title}{{Effect of river training on flood retention of the
  Bavarian Danube}}.
\newblock \bibinfo{journal}{Journal of Hydrology and Hydromechanics}
  \bibinfo{volume}{64}, \bibinfo{pages}{349--356}.
\newblock \DOIprefix\doi{10.1515/johh-2016-0035}.
\bibitem[{{\v{S}}raj et~al.(2016){\v{S}}raj, Viglione, Parajka and
  Bl{\"{o}}schl}]{Sraj2016}
\bibinfo{author}{{\v{S}}raj, M.}, \bibinfo{author}{Viglione, A.},
  \bibinfo{author}{Parajka, J.}, \bibinfo{author}{Bl{\"{o}}schl, G.},
  \bibinfo{year}{2016}.
\newblock \bibinfo{title}{{The influence of non-stationarity in extreme
  hydrological events on flood frequency estimation}}.
\newblock \bibinfo{journal}{Journal of Hydrology and Hydromechanics}
  \bibinfo{volume}{64}, \bibinfo{pages}{426--437}.
\newblock \DOIprefix\doi{10.1515/johh-2016-0032}.
\bibitem[{{Stan Development Team}(2018)}]{StanManual}
\bibinfo{author}{{Stan Development Team}}, \bibinfo{year}{2018}.
\newblock \bibinfo{title}{{Stan Modeling Language Users Guide and Reference
  Manual}version 2.18.0}.
\newblock \bibinfo{note}{\url{http://mc-stan.org}}.
\bibitem[{Statistik~Austria(2017)}]{stataustria}
\bibinfo{author}{Statistik~Austria, S.}, \bibinfo{year}{2017}.
\newblock \bibinfo{title}{Bundesanstalt statistik {\"o}sterreich: Crop
  production 1075 to 2017}.
\newblock \bibinfo{note}{\url{https://www.statistik.at/}, Last accessed:
  2018-01-17}.
\bibitem[{Steirou et~al.(2018)Steirou, Gerlitz, Apel, Sun and
  Merz}]{Steirou2018}
\bibinfo{author}{Steirou, E.}, \bibinfo{author}{Gerlitz, L.},
  \bibinfo{author}{Apel, H.}, \bibinfo{author}{Sun, X.}, \bibinfo{author}{Merz,
  B.}, \bibinfo{year}{2018}.
\newblock \bibinfo{title}{{Do climate-informed extreme value statistics improve
  the estimation of flood probabilities in Europe?}}
\newblock \bibinfo{journal}{Hydrology and Earth System Sciences Discussions} ,
  \bibinfo{pages}{1--23}\DOIprefix\doi{10.5194/hess-2018-428}.
\bibitem[{Szolgayova et~al.(2014)Szolgayova, Parajka, Bl{\"{o}}schl and
  Bucher}]{Szolgayova2014}
\bibinfo{author}{Szolgayova, E.}, \bibinfo{author}{Parajka, J.},
  \bibinfo{author}{Bl{\"{o}}schl, G.}, \bibinfo{author}{Bucher, C.},
  \bibinfo{year}{2014}.
\newblock \bibinfo{title}{{Long term variability of the Danube River flow and
  its relation to precipitation and air temperature}}.
\newblock \bibinfo{journal}{Journal of Hydrology} \bibinfo{volume}{519},
  \bibinfo{pages}{871--880}.
\newblock \DOIprefix\doi{10.1016/j.jhydrol.2014.07.047}.
\bibitem[{{Van der Ploeg} and Schweigert(2001)}]{VanderPloeg2001}
\bibinfo{author}{{Van der Ploeg}, R.}, \bibinfo{author}{Schweigert, P.},
  \bibinfo{year}{2001}.
\newblock \bibinfo{title}{{Elbe river flood peaks and postwar agricultural land
  use in East Germany}}.
\newblock \bibinfo{journal}{Naturwissenschaften} \bibinfo{volume}{88},
  \bibinfo{pages}{522--525}.
\newblock \DOIprefix\doi{10.1007/s00114-001-0271-1}.
\bibitem[{{Van Der Ploeg} et~al.(1999){Van Der Ploeg}, Ehlers and
  Sieker}]{VanDerPloeg1999}
\bibinfo{author}{{Van Der Ploeg}, R.R.}, \bibinfo{author}{Ehlers, W.},
  \bibinfo{author}{Sieker, F.}, \bibinfo{year}{1999}.
\newblock \bibinfo{title}{{Floods and other possible adverse environmental
  effects of meadowland area decline in former West Germany}}.
\newblock \bibinfo{journal}{Naturwissenschaften} \bibinfo{volume}{86},
  \bibinfo{pages}{313--319}.
\newblock \DOIprefix\doi{10.1007/s001140050623}.
\bibitem[{Vehtari et~al.(2017)Vehtari, Gelman and Gabry}]{Vehtari2017}
\bibinfo{author}{Vehtari, A.}, \bibinfo{author}{Gelman, A.},
  \bibinfo{author}{Gabry, J.}, \bibinfo{year}{2017}.
\newblock \bibinfo{title}{{Practical Bayesian model evaluation using
  leave-one-out cross-validation and WAIC}}.
\newblock \bibinfo{journal}{Statistics and Computing} \bibinfo{volume}{27},
  \bibinfo{pages}{1413--1432}.
\newblock \DOIprefix\doi{10.1007/s11222-016-9696-4}.
\bibitem[{Viglione et~al.(2016)Viglione, Merz, {Viet Dung}, Parajka, Nester and
  Bl{\"{o}}schl}]{Viglione2016}
\bibinfo{author}{Viglione, A.}, \bibinfo{author}{Merz, B.},
  \bibinfo{author}{{Viet Dung}, N.}, \bibinfo{author}{Parajka, J.},
  \bibinfo{author}{Nester, T.}, \bibinfo{author}{Bl{\"{o}}schl, G.},
  \bibinfo{year}{2016}.
\newblock \bibinfo{title}{{Attribution of regional flood changes based on
  scaling fingerprints}}.
\newblock \bibinfo{journal}{Water Resources Research} \bibinfo{volume}{52},
  \bibinfo{pages}{5322--5340}.
\newblock \DOIprefix\doi{10.1002/2016WR019036}.
\bibitem[{Viglione et~al.(2013a)Viglione, Merz, Salinas and
  Bl{\"o}schl}]{Viglioneetal2013WRR}
\bibinfo{author}{Viglione, A.}, \bibinfo{author}{Merz, R.},
  \bibinfo{author}{Salinas, J.L.}, \bibinfo{author}{Bl{\"o}schl, G.},
  \bibinfo{year}{2013}a.
\newblock \bibinfo{title}{{Flood frequency hydrology: 3. A Bayesian analysis}}.
\newblock \bibinfo{journal}{Water Resources Research} \bibinfo{volume}{49},
  \bibinfo{pages}{675--692}.
\newblock \DOIprefix\doi{10.1029/2011WR010782}.
\bibitem[{Viglione et~al.(2013b)Viglione, Parajka, Rogger, Salinas, Laaha,
  Sivapalan and Bl{\"{o}}schl}]{Viglione2013}
\bibinfo{author}{Viglione, A.}, \bibinfo{author}{Parajka, J.},
  \bibinfo{author}{Rogger, M.}, \bibinfo{author}{Salinas, J.L.},
  \bibinfo{author}{Laaha, G.}, \bibinfo{author}{Sivapalan, M.},
  \bibinfo{author}{Bl{\"{o}}schl, G.}, \bibinfo{year}{2013}b.
\newblock \bibinfo{title}{{Comparative assessment of predictions in ungauged
  basins - Part 3: Runoff signatures in Austria}}.
\newblock \bibinfo{journal}{Hydrology and Earth System Sciences}
  \bibinfo{volume}{17}, \bibinfo{pages}{2263--2279}.
\newblock \DOIprefix\doi{10.5194/hess-17-2263-2013}.
\bibitem[{Villarini et~al.(2009)Villarini, Smith, Serinaldi, Bales, Bates and
  Krajewski}]{Villarini2009}
\bibinfo{author}{Villarini, G.}, \bibinfo{author}{Smith, J.A.},
  \bibinfo{author}{Serinaldi, F.}, \bibinfo{author}{Bales, J.},
  \bibinfo{author}{Bates, P.D.}, \bibinfo{author}{Krajewski, W.F.},
  \bibinfo{year}{2009}.
\newblock \bibinfo{title}{{Flood frequency analysis for nonstationary annual
  peak records in an urban drainage basin}}.
\newblock \bibinfo{journal}{Advances in Water Resources} \bibinfo{volume}{32},
  \bibinfo{pages}{1255--1266}.
\newblock \DOIprefix\doi{10.1016/j.advwatres.2009.05.003}.
\bibitem[{Volpi et~al.(2018)Volpi, Di~Lazzaro, Bertola, Viglione and
  Fiori}]{Volpi2018}
\bibinfo{author}{Volpi, E.}, \bibinfo{author}{Di~Lazzaro, M.},
  \bibinfo{author}{Bertola, M.}, \bibinfo{author}{Viglione, A.},
  \bibinfo{author}{Fiori, A.}, \bibinfo{year}{2018}.
\newblock \bibinfo{title}{{Reservoir effects on flood peak discharge at the
  catchments scale}}.
\newblock \bibinfo{journal}{Water Resources Research} .
\bibitem[{Vorogushyn and Merz(2013)}]{Vorogushyn2013}
\bibinfo{author}{Vorogushyn, S.}, \bibinfo{author}{Merz, B.},
  \bibinfo{year}{2013}.
\newblock \bibinfo{title}{{Flood trends along the Rhine: the role of river
  training}}.
\newblock \bibinfo{journal}{Hydrology and Earth System Sciences}
  \bibinfo{volume}{17}, \bibinfo{pages}{3871--3884}.
\newblock \DOIprefix\doi{10.5194/hess-17-3871-2013}.
\bibitem[{Watanabe(2010)}]{Watanabe2010}
\bibinfo{author}{Watanabe, S.}, \bibinfo{year}{2010}.
\newblock \bibinfo{title}{Asymptotic equivalence of bayes cross validation and
  widely applicable information criterion in singular learning theory}.
\newblock \bibinfo{journal}{J. Mach. Learn. Res.} \bibinfo{volume}{11},
  \bibinfo{pages}{3571--3594}.

\end{thebibliography}





\end{document}